\documentclass[aps,prb,twocolumn,groupedaddress,showpacs,amssymb,amsmath,letterpaper]{revtex4}
\usepackage{graphicx}
\usepackage[final,dvips]{epsfig}
\usepackage{color}
\usepackage{dcolumn}
\usepackage{bm}
\usepackage{array}
\usepackage{amssymb}

\definecolor{Green}{rgb}{0.2,0.96,0.2}
\definecolor{Remarks}{rgb}{1,0.3,0.3}
\definecolor{Extra}{rgb}{0.2,0.2,1}
\definecolor{Blue}{rgb}{0.2,0.3,1}
\definecolor{Black}{rgb}{0,0,0}
\definecolor{CC}{rgb}{0.2,0.75,0.2}

\newcommand{\bi}{\begin{itemize}}
\newcommand{\ei}{\end{itemize}}
\newcommand{\be}{\begin{equation}}
\newcommand{\ee}{\end{equation}}
\newcommand{\ba}{\begin{eqnarray}}
\newcommand{\ea}{\end{eqnarray}}

\newcommand{\upa}{\uparrow}
\newcommand{\dna}{\downarrow}

\newcommand{\COMMENTED}[1]{}

\newcommand{\ob}[1]{\langle #1\rangle}

\newcommand{\bfc}{\mathbf{c}}

\newcommand{\bfe}{\mathbf{e}}

\newcommand{\bfk}{\mathbf{k}}

\newcommand{\bfq}{\mathbf{q}}
\newcommand{\bfr}{\mathbf{r}}

\newcommand{\bfG}{\mathbf{G}}

\newcommand{\bfL}{\mathbf{L}}

\newcommand{\bfQ}{\mathbf{Q}}
\newcommand{\bfR}{\mathbf{R}}
\newcommand{\bfS}{\mathbf{S}}

\newcommand{\calH}{{\cal H}}

\begin{document}

\title{Magnetic order in the repulsive Fermi-Hubbard model in three-dimensions and 
the crossover to two-dimensions}

\author{Jie Xu, Simone Chiesa, Eric J. Walter, Shiwei Zhang}
\affiliation{Department of Physics, College of William and Mary, Williamsburg, VA 23187, USA}

\begin{abstract}
Systems of fermions described by the three-dimensional (3D) repulsive Hubbard model
on a cubic lattice have recently attracted considerable attention due to
their possible experimental realization via cold atoms in an optical lattice.
Because analytical and numerical results are limited away from half-filling,
we study the ground state of the doped system from weak to intermediate interaction 
strengths within the generalized Hartree-Fock approximation.
The exact solution to the self-consistent-field equations in the thermodynamic
limit is obtained and the ground state is shown
to exhibit antiferromagnetic order and incommensurate spin-density waves (SDW). 
At low interaction strengths, the SDW has unidirectional character with a leading wave-vector 
along the $\langle100\rangle$-direction, and the system is metallic.
As the interaction increases, the system undergoes a simultaneous structural and metal-to-insulator 
transition to a unidirectional SDW state 
along the $\langle111\rangle$-direction, with a different wavelength.
We systematically determine the real- and momentum-space properties of these states.
The crossover from 3D to two-dimensions (2D) is then studied by
varying the inter-plane hopping amplitude,
which can be experimentally realized by tuning the
distance between a stack of square-lattice layers. 
Detailed comparisons are made between the exact numerical results and predictions from the pairing model,  
a variational {\em ansatz} based on the pairing of spins in the 
vicinity of the 
Fermi surface. Most of the numerical results can be understood quantitatively from
the ansatz, which provides a simple picture for the nature of the SDW states.
\end{abstract}

\pacs{75.30.Fv, 71.15.Ap, 71.45.Lr, 71.10.Fd, 75.10.Lp, 75.50.Ee}

\maketitle

\section{Introduction}
\label{sec:intro}

Over the past several years, optical lattices have become
an increasingly powerful tool for emulating many systems in condensed matter physics
\cite{Morsch2006, Lewenstein2007, Bloch2008, Esslinger2010}.
An optical lattice can provide exceptionally clean access
to a variety of model many-body Hamiltonians in which parameters can be systematically tuned
and controlled. Thus, they make possible quantitative experimental study of the properties of interacting electron 
models, which have proven extremely challenging for analytic and numerical approaches alone.
The combination of these approaches presents unprecedented opportunities for 
improving our understanding of interacting electron systems, by
testing theoretical concepts and increasing the accuracy and predictive power of 
numerical approaches via comparison with experiment.

The one-band Hubbard model is one of the most fundamental models in condensed matter physics.
It has been widely studied in two dimensions (2D) \cite{Su1988, Zaanen1989,
Schulz1990, Machida1990, Singh1990, Littlewood1991, Yang1991, Ichimura1992, Scalapino1994,
Zhang1997, Zitzler2002, Maier2005, Capone2006, Aimi2007, Chang2010, Xu2011}
as the simplest model for the Cu-O plane in cuprate superconductors.
For the three-dimensional (3D) Hubbard model, however, considerably less is known
from both theoretical and experimental sides.
Optical lattices play, in this respect, a particularly fundamental role as they allow 
for a clean experimental realization of the 3D model and offer the interesting
possibility of tuning the hopping parameter along one direction, $t_\perp$,
thereby allowing a systematic study of the evolution of
properties as the system crosses over from 3D to 2D.

Thanks to advances in the ability to directly cool atoms in optical lattices\cite{Mathy2012},
experiments are nearing the realization of phases with magnetic order. 
It is thus particularly important and timely to understand such phases in 
the Hubbard model.
Somewhat surprisingly, apart from half-filling (one particle per site), 
which displays antiferromagnetic (AFM) order,
the nature of the magnetic properties in the
3D Hubbard model has not been
characterized, even at the mean-field level.
In this work,
we study the magnetic properties in the ground states of the 3D Hubbard model 
and in the crossover regime, using generalized Hartree-Fock (HF) theory.
It is shown that the system has a tendency to form unidirectional spin-density
wave (SDW) states with AFM order and a modulating wave along
either the $\langle100\rangle$- (at low $U/t$)
or the $\langle111\rangle$-direction (at higher $U/t$).
We examine the evolution of the SDW wavelength in the full mean-field solution 
as $U$, density and $t_\perp$ vary and characterize the ground state
by its properties in real and momentum space.

Despite the simple nature of the mean-field approach,
the determination of the correct equilibrium properties
in these models is not straightforward \cite{Xu2011, Zhang2008}.
The main challenge lies in finding an unbiased strategy
to determine the leading wave-vector(s) characterizing
the spatial dependence of the order parameter. Calculations are performed 
in a real space simulation cell
and most choices of the cell
will return solutions that are biased by finite-size effects.
This is often further complicated by shell effects and sensitivity of the 
solution to the 
topology of the Fermi surface, which often lead to local minimum solutions. 

To overcome the difficulties, it is necessary to move to larger and larger cells
and gain insights from the evolution of the corresponding solutions.
This line of attack has become increasingly possible because of
the dramatic increase in computing power and continuous algorithmic progress.
In the present work, we combine such an approach with more targeted searches
to obtain the global minimum solution of the self-consistent-field (SCF) equations in the thermodynamic limit.
Furthermore, we show how the numerical results can be understood
by a variational ansatz based on the pairing of spins in the vicinity of the
Fermi surface. Detailed comparisons are made between the direct numerical 
solutions and the pairing ansatz predictions. The excellent agreement helps to 
provide a simple, predictive picture for the properties of the SDW states.

The mean-field approach is often the starting point in the study of
strongly interacting systems such as the Hubbard model. Although the 
approximations involved can lead to significant errors, mean-field theory 
often provides insights into qualitative aspects
of the behavior of many-body systems.
Moreover, comparisons with quantum Monte Carlo results \cite{Chang2010} have
shown \cite{Xu2011} that, in 2D, the mean-field solution captures the 
basic physics of SDW states at intermediate interaction strengths, and 
provides a good qualitative (or even quantitative in some aspects) description
of the magnetic correlations in the true ground state.
Because it is reasonable to expect a similar level of accuracy for the models presently considered, 
we expect that our findings will provide guidance to many-body approaches and experimental 
studies alike.

We have limited our study to $U\lesssim 6t$, 
where the mean-field approach can be expected to offer useful insight.
Below we will discuss the mean-field predictions and their implications (and the caveats) 
on the true many-body states drawing from the comparison in 2D \cite{Chang2010,Xu2011}.
Clearly, the form of generalized mean-field theory considered in this work will not 
capture exotic instabilities, such as unconventional
pairing order. Indeed, as $U$ increases, it will become increasingly inadequate for
magnetic properties as well.

The remainder of the paper is organized as follows.  In
Sec.~\ref{sec:background}, we introduce the Hamiltonian,
and briefly outline some of its basic properties to facilitate the ensuing discussion.
In Sec.~\ref{sec:method}, we summarize
the strategies used to solve the mean-field equations.
Results for the 3D model are presented in Sec.~\ref{sec:3D};
numerical results for the $\langle100\rangle$- and the $\langle111\rangle$-SDW
are followed by a discussion where the pairing ansatz is first introduced
and then used to help understand the numerical findings.
The dimensional crossover results are then presented in Sec.~\ref{sec:crossover},
followed again by a discussion based on insights from the pairing ansatz.
We conclude in Sec.~\ref{sec:disc&conclu}. 

\section{Background}
\label{sec:background}

Given our goal to study both the 3D case and 
the crossover from 3D to 2D, 
it is most convenient to define the 3D Hubbard Hamiltonian
as a stack of square-lattice planes. We will use $\bfr\equiv(x,y)$ to denote in-plane coordinates
and $z$ to label the planes.
With this convention the Hubbard Hamiltonian reads
\ba
  \calH &=& -t \sum_{\langle\bfr\bfr'\rangle, z, \sigma}
                     \left( c_{\bfr z \sigma}^\dagger c_{\bfr'z \sigma} +
                              c_{\bfr'z \sigma}^\dagger c_{\bfr z\sigma} \right) \nonumber \\
               && -t_\perp \sum_{\bfr, \langle zz' \rangle, \sigma}
                     \left( c_{\bfr z \sigma}^\dagger c_{\bfr z' \sigma} +
                              c_{\bfr z' \sigma}^\dagger c_{\bfr z\sigma} \right) \nonumber \\
               && + U\sum_{\bfr, z}n_{\bfr z \uparrow} n_{\bfr z \downarrow},
\label{eq:def_ham}
\ea
where the operator $c_{\bfr z \sigma}^\dagger$ ($c_{\bfr z \sigma}$) creates (annihilates)
a particle with spin $\sigma$ ($\sigma=\uparrow, \downarrow$) at site $(\bfr, z)$ and
$n_{\bfr z \sigma}$ is the corresponding number operator.
The hopping amplitude $t$ is between nearest neighbor sites within a plane
(denoted by $\langle\bfr\bfr'\rangle$ in the summation), 
$t_\perp$ is the inter-plane hopping amplitude between nearest neighbor sites belonging to 
different  planes
(denoted by $\langle z z' \rangle$ in the summation),
and $U>0$ is the on-site interacting strength.
Throughout this work, energy is quoted in units of $t$ and we set $t=1$.
The Hamiltonian in Eq.~(\ref{eq:def_ham}) describes the 3D cubic Hubbard model when $t_\perp=1$,
the crossover between the square and cubic lattices when $0<t_\perp<1$
and 
a stack of decoupled 2D Hubbard planes when $t_\perp=0$.
Only unpolarized systems are considered, {\em i.e.}, the average densities of the two spin species 
are kept equal: 
$n_\uparrow=n_\downarrow$.
The nature of the ground state is thus characterized by three parameters:
the inter-plane hopping amplitude $t_\perp$, the on-site repulsion $U$ and the 
doping (hole density)
\begin{equation}
h\equiv 1-(n_\uparrow+n_\downarrow).
\label{eq:def_doping}
\end{equation}
The particle-hole transformation,
$c^\dagger_{\bfr z \sigma} \rightarrow (-1)^{x+y+z} c_{\bfr z \sigma}$,
maps the $h<0$ sector into the $h>0$ one, regardless of the value of $t_\perp$ or $U$,
and we therefore confine our study to $h > 0$.

\begin{figure}
\includegraphics[width=\columnwidth]{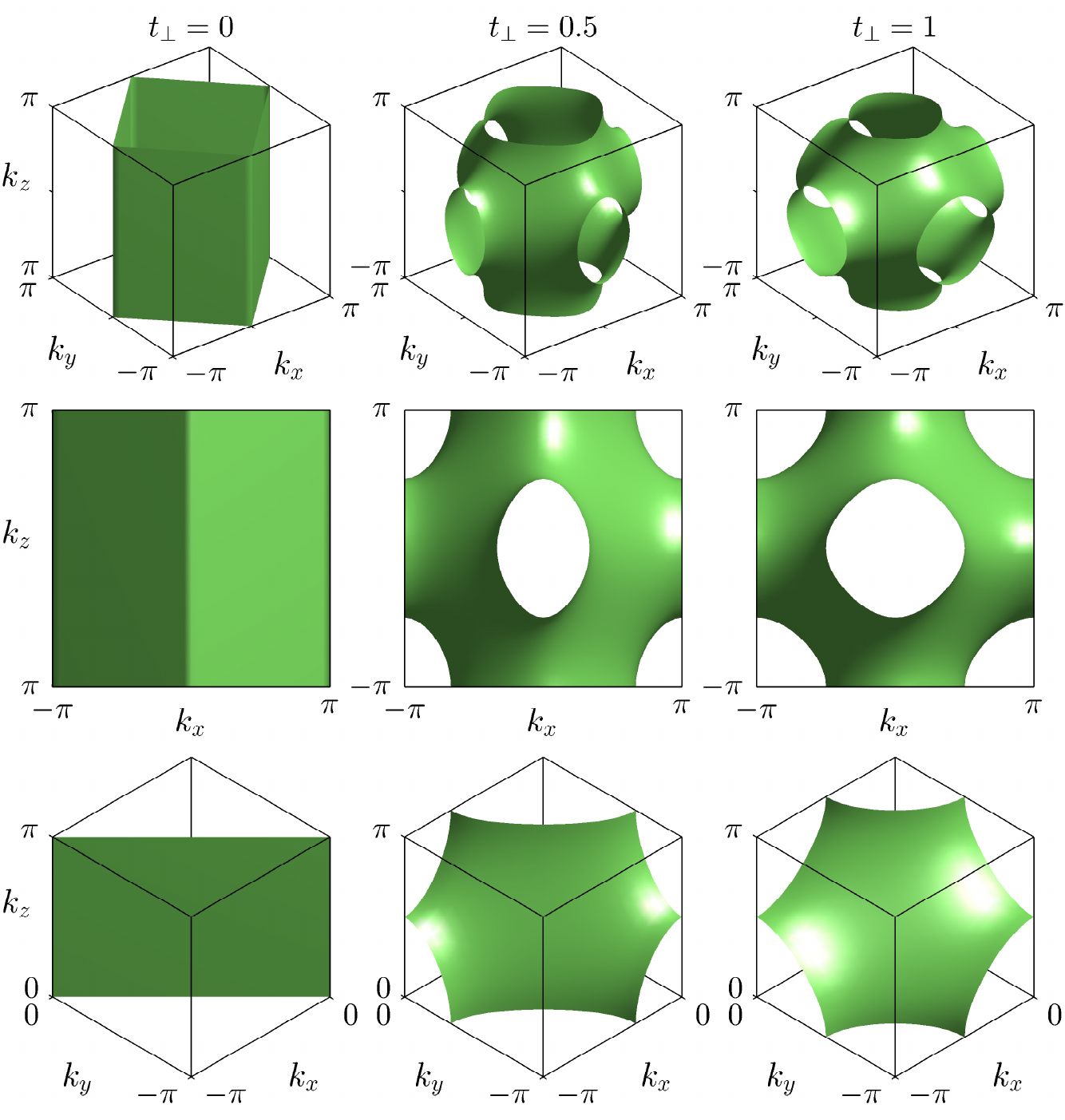}
\caption{
(Color online)
Non-interacting half-filled Fermi surface from different view angles:
3D (top), along $[010]$ (middle), along $[11\bar1]$ (bottom).
From left to right, the columns are for
$t_\perp=0$, $0.5$ and $1$, respectively.
Note that only the surface in one octant is shown in the bottom row.
Perfect nesting via $\bfQ=(\pm\pi, \pm\pi, \pm\pi)$
holds for any $t_\perp$ at half-filling.
}
\label{fig:halffillingFS}
\end{figure}

At half-filling, $h=0$,
the non-interacting Fermi surface is given by
$-2(\cos k_x+\cos k_y+t_\perp\cos k_z)=0$.
Despite the lack of symmetry
between the $z$- and the $\bfr$-directions
for any $t_\perp \ne 1$,
perfect nesting  via 
$\bfQ\equiv(\pm\pi, \pm\pi, \pm\pi)$ remains throughout the whole surface, and causes
an AFM instability for any $t_\perp\ne 0$ and arbitrary 
small $U$ values. 
The evolution of the non-interacting half-filled Fermi surface as $t_\perp$ varies
is shown in Fig.~\ref{fig:halffillingFS}. In the first column, 
representing the 2D limit, the Fermi surface has no dependence on $k_z$ and
any wave-vector of the form $(\pm\pi,\pm\pi,q)$ is perfectly nested on it.
The arbitrariness of $q$ is reflected in the complete lack of correlation between the 
$\bfr$-planes.
The  large nesting degeneracy is abruptly interrupted as soon as $t_\perp \ne 0$,
and $\bfQ$ remains the only nesting vector as the system evolves from the 2D limit
toward 3D.
The middle and bottom rows illustrate projections of the Fermi surface
along the $\langle100\rangle$- and $\langle111\rangle$-directions;
as we shall see in Sec.'s~\ref{sec:3D} and \ref{sec:crossover}, the projected area of the Fermi surface
plays a central role in determining the character of the SDW in the proximity of $h=0$.

\section{Method}
\label{sec:method}

The following mean-field formalism in real space is used in this work.
A simulation cell 
of $N$ sites is defined by three vectors,
$\bfL_1$, $\bfL_2$ and $\bfL_3$, whose components are integers.
Bloch states are then introduced as
\begin{equation}
c_\beta (\bfk) \propto \sum_{\bfL} c_{\beta+\bfL} \exp\big[i\bfk\cdot \bfL\big],
\end{equation}
where $\bfL$ are vectors of the form $\bfL=n_1 \bfL_1 + n_2 \bfL_2 + n_3 \bfL_3$,
$\bfk$ is a reciprocal lattice vector that is free to vary 
within the first Brillouin zone (BZ) defined by the $\bfL_i$'s, and
$\beta$ labels sites inside the simulation cell. 
Using these states, the mean-field Hamiltonian can be decoupled
into a sum of $\bfk$-dependent pieces, $H_0 = \sum_\bfk H_0(\bfk)$,
with each piece of the form
\begin{equation}
\begin{split}
H_0(\bfk)= [\bfc^\dagger_\uparrow \bfc^\dagger_\downarrow] 
\left[ \begin{array}{cc}
\mathbb{H}_\uparrow(\bfk) & \mathbb{S}^-  \\
\mathbb{S}^+ & \mathbb{H}_\downarrow(\bfk-\bfG)  \end{array} \right]
[\bfc_\uparrow \bfc_\downarrow]^T,
\end{split}
\label{eq:MFH}
\end{equation}
where $\mathbf{c}_\uparrow$ ($\mathbf{c}_\downarrow$) represents
a 
row of operators $c_{\beta\uparrow}(\bfk)$ ($c_{\beta\downarrow}(\bfk-\bfG)$)
with index $\beta$ running through the $N$ sites of the cell. 
A non-zero value of $\bfG$ causes the spin densities at $\beta$ and $\beta+\bfL_i$
to be related via a rotation by $\bfG\cdot \bfL_i$ around the $z$-axis. 
Charge and spin densities along $z$-direction
obey periodic boundary conditions.
$\mathbb{H}$ and $\mathbb{S}^\pm$ are $N\times N$
matrices with elements
\begin{equation}
\begin{split}
[\mathbb{H}_\sigma(\bfk)]_{\beta\gamma} & = -t_{\beta\gamma}(\bfk) + \delta_{\beta\gamma} (U D_{\beta\sigma} - \mu), \\
[\mathbb{S^\pm}(\bfk)]_{\beta\gamma}    & = U \delta_{\beta\gamma} S^\pm_\beta,
\end{split}
\end{equation}
where $t_{\beta\gamma}(\bfk)=\sum_\bfL \exp(i\bfk\cdot\bfL) t_{\beta,\gamma+\bfL}$, 
and $D_{\beta\sigma}$, $S^\pm_\beta$ and $\mu$ are determined by the
requirement that the free energy $F = \langle H\rangle_0  - T S_0$
is a minimum for the targeted average density $n=n_\uparrow+n_\downarrow$.
This amounts to the following SCF (gap) equations
\begin{equation}
\begin{split}
D_{\beta,-\sigma} =&  \frac{1}{(2\pi)^3} \int d\bfk \langle c^\dagger_{\beta\sigma}(\bfk)c_{\beta\sigma}(\bfk) \rangle_0 \\
S^\pm_\beta        =& -\frac{1}{(2\pi)^3} \int d\bfk \langle c^\dagger_{\beta,\pm\sigma}(\bfk)c_{\beta,\mp\sigma}(\bfk) \rangle_0\\
n  =&  \frac{1}{N(2\pi)^3}\sum_{\beta,\sigma} \int d\bfk \langle c^\dagger_{\beta\sigma}(\bfk)c_{\beta\sigma}(\bfk)\rangle.
\end{split}
\label{eq:gapEq}
\end{equation}

To locate the ground state we proceed with two complementary approaches.
In the \emph{first approach} we select the $\bfL_i$'s so that
they span a large supercell containing 
$\mathcal{O}(5000)$ sites.
A twisted boundary condition \cite{Poilblanc1991,Lin2001} is applied, 
namely, using a single randomly selected $k$-point  in place of the integrals in 
 Eq.~(\ref{eq:gapEq}).
The iterative process is started with various initial states, including random ones,
and multiple annealing cycles are performed. In each cycle a random perturbation 
(whose strength can be controlled) is applied to a converged solution
and the self-consistent process is repeated. Separate calculations for different $k$-points
are done to check for consistency.

Once an understanding of the character of the ground state is gained,
we use a \emph{second approach} to target the specific family of states 
compatible with the results of the random search.
For instance, suppose the random search finds a unidirectional 
SDW at small $U$ values with wave-vector along the $\langle100\rangle$-direction.
We then choose a cluster of $\bfL_1=(L,0,0)$, $\bfL_2=(0,1,0)$ and $\bfL_3=(0,0,1)$
with $\bfG=\pi((-1)^{L+2l},1,1)$, where $l$ is the number of 
oscillations of the order parameter,  chosen to be an integer or half an integer.
For a given set of the three parameters ($t_\perp$, $U$, $h$),
$L$ is finely scanned (with $L$ on the order of $50$ and step size of $1$)
until the energy minimum is found.
A large number of $k$-points is used
(on the order of $100$ in the two short directions and a few in the other)
so that the character and properties of the targeted states can be accurately determined.
This approach allows us to study different forms of SDW and long wavelength modes
without increasing the computational cost.

In our study,  we mix the two numerical approaches as needed and use them in complementary ways. For example, 
comparison of energies among several families of SDW is made with the second 
approach. To confirm the correctness of the ground state, 
the solutions are then checked against different initial states and annealing procedures
using the first approach on a supercell commensurate with the optimal wavelength.

Various observables are computed to characterize
the converged solutions.
The local charge density $\rho$ and the local order parameter, identified as
the local staggered magnetization $m$, are defined as
\begin{eqnarray}
  \rho(\bfR) &\equiv& \ob{ n_{\bfr z \upa} }+\ob{ n_{\bfr z \dna} }, \\
      m(\bfR) &\equiv& (-1)^{x+y+z}\left(\ob{ n_{\bfr z \upa} }-\ob{ n_{\bfr z \dna} }\right),
\label{eq:def_CDSD}
\end{eqnarray}
and used to characterize the state in real space (here $\bfR\equiv(\bfr,z)$). 
Since all the minimum energy solutions we find are unidirectional 
spin/charge density waves (SDW/CDW), it is natural to characterize them by their
modulation wavelength along a relevant Cartesian axis, $\lambda_{\rm SDW/CDW}$, 
defined by the leading component of the  Fourier transform of $\rho$ and $m$, respectively.
The minima in the CDW are found to coincide
with nodes in the SDW,
thus $2\lambda_\mathrm{CDW}=\lambda_\mathrm{SDW}$. 
Below we will sometimes discuss our results in terms of a single wavelength $\lambda\equiv\lambda_\mathrm{CDW}$,
which can also be identified as the distance between two consecutive nodes of the order parameter.
When we refer to the direction of the modulation, we will use
$\langle100\rangle$ to denote symmetry-equivalent  $[100]$-directions, and similarly for 
$[110]$ and $[111]$.

\begin{figure}[t]
\includegraphics[width=\columnwidth]{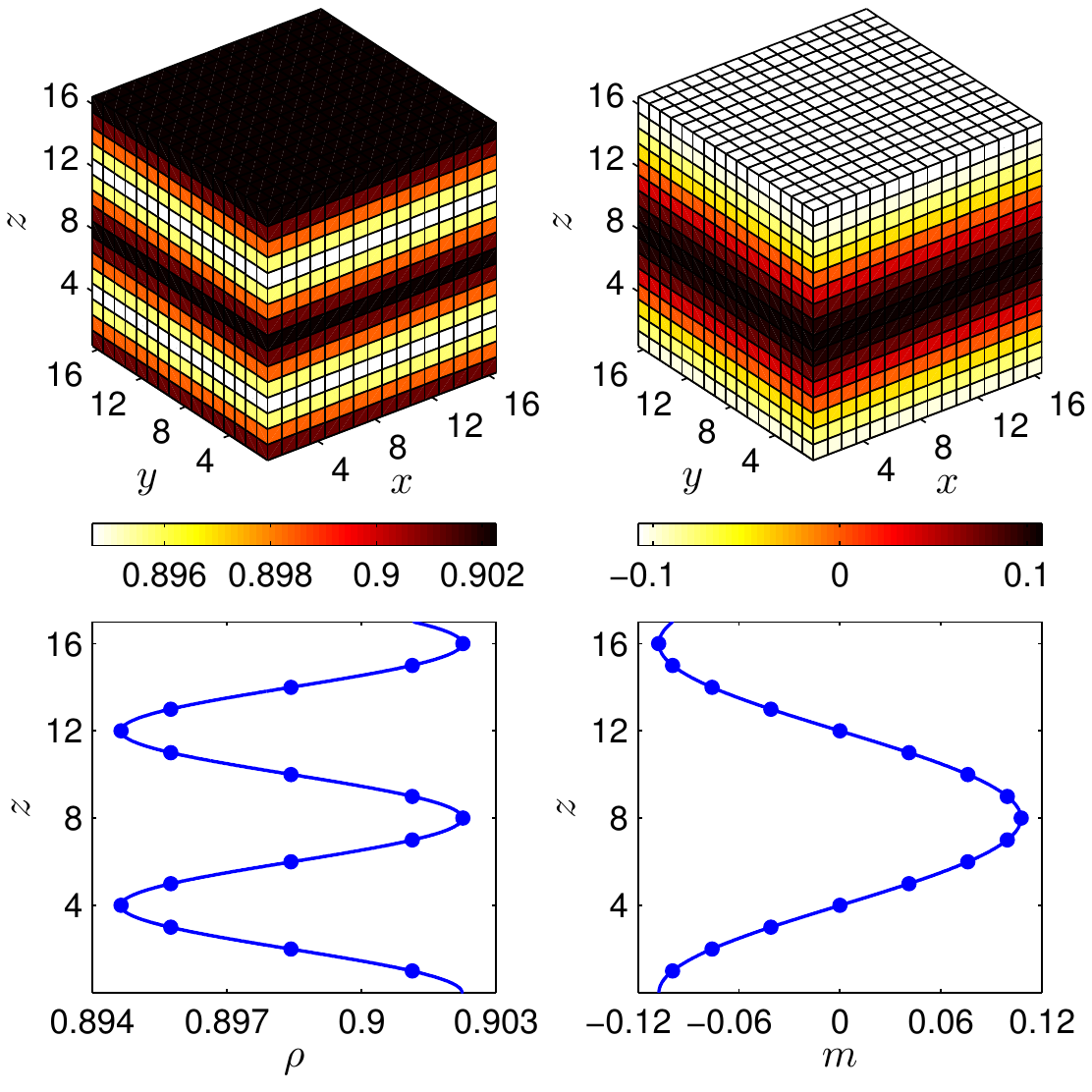}
\caption{
(Color online) Charge density
$\rho$ (left) and order parameter $m$ (right) of the solution for the 3D Hubbard model.
Shown is a $16 \times 16 \times 16$ supercell, with $h=13/128$ at $U=2.5$.
A linear wave is seen along the $z$-direction, with uniform AFM
order in the $xy$-plane.
The bottom panel shows a line cut along $z$-direction, with dots the actual data and the line a  
 sinusoidal fit.
}
\label{fig:rho_m_U25_3D}
\end{figure}

To characterize the system in momentum space, we use
the momentum distribution $n_{\bfk}$ and the momentum-resolved 
single-particle spectral function $A_{\bfk}(\omega)$,
defined as
\begin{eqnarray}
&n_{\bfk \sigma}=\langle c_{\bfk\sigma }^\dagger c_{\bfk\sigma}\rangle,\\
&A_{\bfk \sigma}(\omega)=\frac{1}{\pi}\mathrm{Im}\langle c_{\bfk\sigma} (\omega - H + E_0 -i\eta)^{-1}  c_{\bfk\sigma}^\dagger \rangle
\end{eqnarray}
with $c_{\bfk\sigma}\propto\sum_\bfR \exp(-i\bfk\cdot\bfR) c_{\bfR\sigma}$.
We use
$n_{\bfk}$ to compare the converged mean-field solution with
the pairing ansatz prediction 
and $A_\bfk$ to characterize the Fermi surface of the ordered phase.

\section{3D Results}
\label{sec:3D}

\subsection{SDW correlation in the $\langle100\rangle$-direction}
\label{ssec:weak_3D}

At half-filling, the existence of perfect nesting allows an AFM solution for any $U>0$.
Away from half-filling, perfect
nesting ceases to exist and a finite critical value of the interaction
is needed to cause the onset of order. The critical value $U_c$ depends on $h$.
Using the first approach described in Sec.~\ref{sec:method}, we have
determined that, just above $U_c$, the
ground state of the system
is an SDW with modulation along the $\langle100\rangle$-direction. 
Figure \ref{fig:rho_m_U25_3D} illustrates the spatial dependence of
$\rho$  and $m$ 
in a $16 \times 16 \times 16$ supercell at $h=13/128\simeq0.10$ and $U=2.5$.
The SDW is characterized by a single wave-vector and $\lambda_{\langle100\rangle}=8$.
The amplitude of the SDW is $\simeq 0.1$, roughly thirty times larger than that of the CDW. 
The simple form of the order found for $m(\bfR)$ is indicative
of the proximity of $U$ to $U_c$.
All of the observations above are 
consistent with the pairing model, as discussed below in 
Sec.~\ref{ssec:pairing}.

\begin{figure}[t]
\includegraphics[width=\columnwidth]{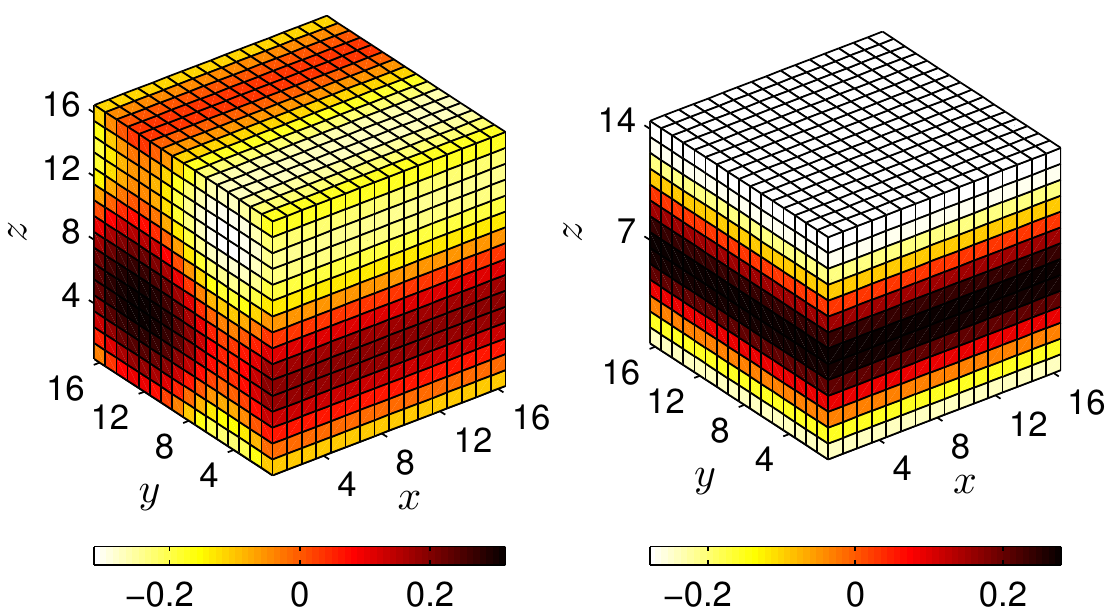}
\caption{
(Color online) Spatial dependence of the order parameter $m$ for $h=13/128$,
same as in Fig.~\ref{fig:rho_m_U25_3D}, but with $U=2.9$,
on a $16 \times 16 \times 16$ supercell (left) and a $16 \times 16 \times 14$ supercell (right).
Uniform AFM order in the $xy$-plane disappears for $L_z=16$,
but linear SDW along $z$-direction is seen again on the right with $L_z=14$.
}
\label{fig:m_U29_3D}
\end{figure}

\begin{figure}
\includegraphics[width=\columnwidth]{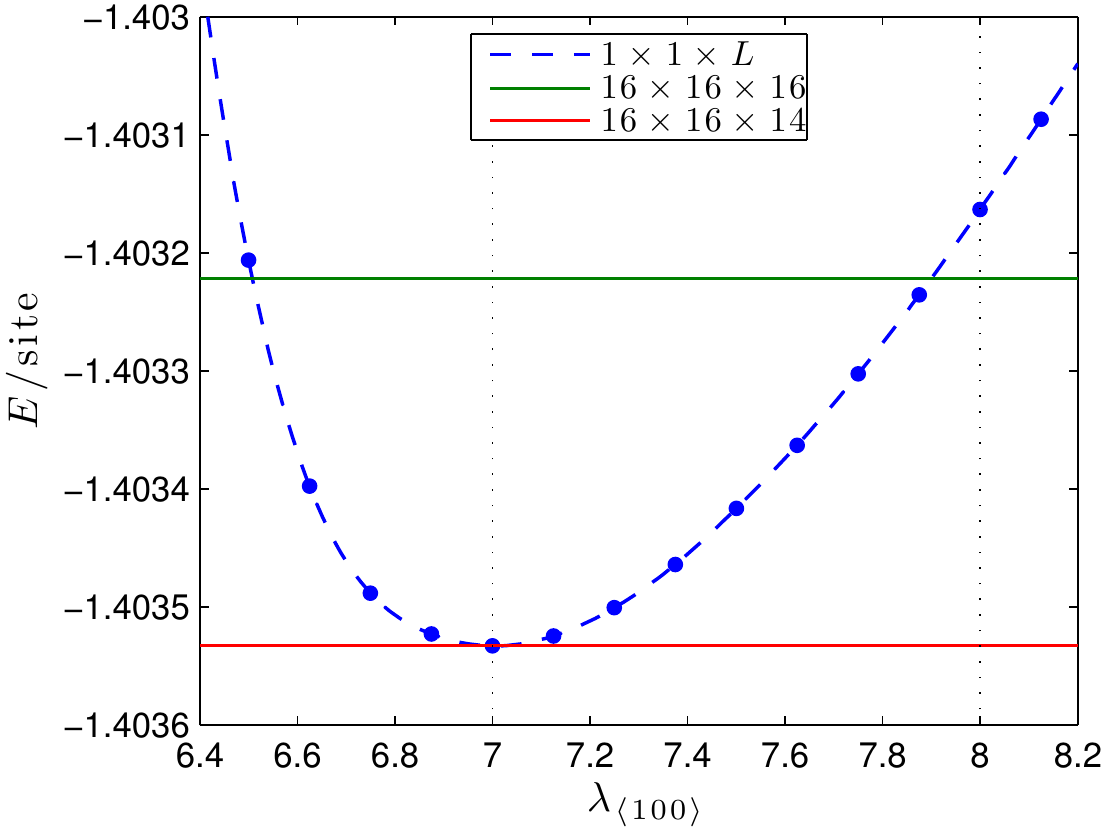}
\caption{
(Color online) Energy of $\langle100\rangle$-SDW (blue) vs. $\lambda_{\langle100\rangle}$
for a system of $h=13/128$ and $U=2.9$. Horizontal lines are 
the energies of the calculations shown in Fig.~\ref{fig:m_U29_3D}.
The minimum of $\langle100\rangle$-SDW is reached when $\lambda_{\langle100\rangle}=7$.
The state in the left panel of Fig.~\ref{fig:m_U29_3D} leads to an energy 
higher than the minimum but lower than the energy with $\lambda_{\langle100\rangle}=8$.
}
\label{fig:E_n08984375_U29}
\end{figure}

We next examine the evolution of $\lambda_{\langle100\rangle}$ as the interaction strength changes.
Keeping $h$ and the simulation cell unchanged and increasing $U$ from 2.5 to 2.9, 
the random search returns the state displayed on
the left panel of Fig.~\ref{fig:m_U29_3D}, which suggests that
a more complicated type of order is seemingly settling in.
We apply our second approach, using a dense $k$-point grid and a $1\times1\times L$ simulation cell,
to search for the optimal wavelength. 
We use large $L$ (containing about 8 nodes in the cell)
and vary
its value until an energy minimum is found. 
Figure \ref{fig:E_n08984375_U29} shows the result
of such minimization in terms of $\lambda_{\langle100\rangle}$;
the minimum occurs when $\lambda_{\langle100\rangle}=7$,
indicating that the $16 \times 16\times 16$ supercell
is not commensurate with the wavelength of the minimum energy SDW state
and that the pattern in the left panel of  Fig.~\ref{fig:m_U29_3D}  is a result of frustration from an incommensurate supercell size.
We next return to our first approach, and 
perform a new mean-field calculation, with {\em random} 
initial guess and annealing, on a $16 \times 16\times 14$ supercell,
a size which is commensurate with the wavelength of the minimum energy solution.
And indeed we find the predicted state with $\lambda_{\langle100\rangle}=7$ correctly reproduced (right panel in Fig.~\ref{fig:m_U29_3D}).

In Fig.~\ref{fig:E_n08984375_U29}, we report the energies of the two large supercell calculations
of Fig.~\ref{fig:m_U29_3D},
to verify that the energy obtained in the $16 \times 16\times 14$ supercell
is correctly reproduced by the $1\times1\times L$ cluster search with
$\lambda_{\langle100\rangle}=7$.
The energy of the $16 \times 16\times 16$ calculation, on the other hand, falls between
those of $\lambda_{\langle100\rangle}=7$ and 8.
This gives a clear illustration of the characteristics of the two types of approaches.
The supercell being incommensurate 
prevents the solution from collapsing onto the lowest energy SDW 
state of $\lambda_{\langle100\rangle}=7$. 
The self-consistent solution in a large supercell then 
finds a different pattern that corresponds to the true ground state
compatible with the imposed constraint.
The energy of this state, computed by fixing the density 
and converging the value using a dense $k$-point mesh, is higher than the global minimum, 
but lower than that of the 
SDW state with an imposed wavelength $\lambda_{\langle100\rangle}=8$.

\begin{figure}
\includegraphics[width=\columnwidth]{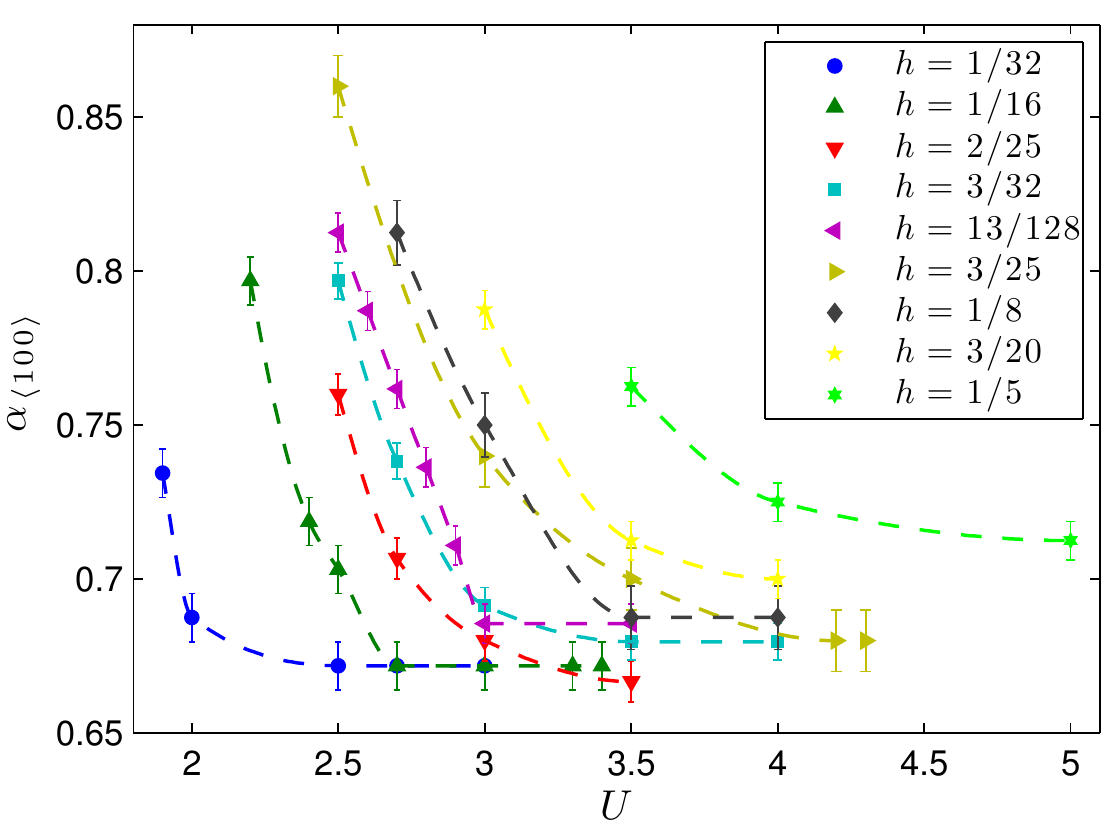}
\caption{
(Color online) Characteristic wavelength  
as a function of $U$ at various doping.
$\alpha_{\langle100\rangle}$ gives the 
modulation wavelength, $\lambda_{\langle 100\rangle}$, in units of $1/h$. 
As $U$ is increased, the value of $\alpha_{\langle100\rangle}$
converges to approximately 2/3 at small $h$ (slightly larger at larger $h$).
}
\label{fig:alpha_3D}
\end{figure}

We proceed to determine the exact dependence of the wavelengths on $h$ and $U$
by explicit solutions of the SCF equations in $1\times 1\times L$ clusters.
Verifications of the results are done on large supercells
whose sizes are commensurate with the wavelengths.
Our results are summarized in Fig.~\ref{fig:alpha_3D},
with $\alpha_{\langle100\rangle}$ defined as
\begin{eqnarray}
  \alpha_{\langle100\rangle} = h \lambda_{\langle100\rangle}.
\label{eq:def_alpha_CDW} 
\end{eqnarray}
When the doping is small, the 
wavelength of the modulation is proportional to $1/h$, with 
$\alpha_{\langle100\rangle}$ 
almost independent of $U$ and roughly equal to 2/3.
For larger $h$, $\alpha_{\langle100\rangle}$ converges to a slightly larger value.
There is a general trend of an increase of $\alpha_{\langle100\rangle}$
as $U$ approaches $U_c$ from above.
We will be able to rationalize these trends within the pairing model
in Sec.~\ref{ssec:pairing}.

\begin{figure}
\includegraphics[width=\columnwidth]{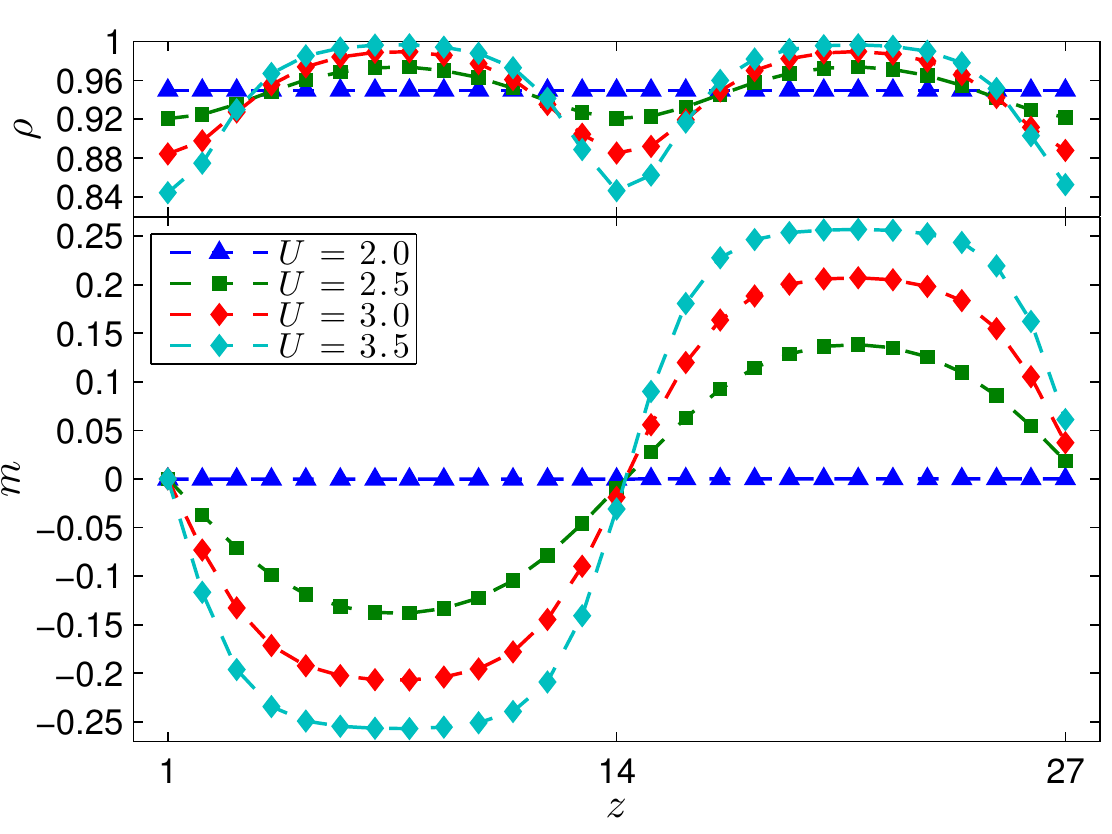}
\caption{
(Color online) Charge density $\rho$
(top) and order parameter $m$
(bottom) 
vs.~$U$.
The system has doping $h=0.05$. 
Each curve is a 1D cut along $z$-direction, the direction of the modulation. 
Beyond $U_{\rm c}$, the $\langle100\rangle$-SDW/CDW amplitudes increase 
with $U$ 
and the solution evolves from a sinusoidal wave to domain walls.
The CDW amplitude is much weaker than that of the SDW.
}
\label{fig:rho_m_n095_3D}
\end{figure}

The evolution of the
properties of the 3D SDW state with interaction $U$ is similar 
to what is observed in 2D. Figure \ref{fig:rho_m_n095_3D} shows 1D cuts of $m$ and $\rho$
in the $z$-direction, at $h=0.05$ and $U$ values such that $\alpha_{\langle100\rangle}$ has saturated to $\sim 2/3$.
The figure illustrates the existence of $U_c$ and
the increase of the SDW and CDW amplitudes with $U$.
It also shows the crossover from a regime where the order parameter has a smooth sinusoidal modulation and the holes are delocalized,  
to one where it is characterized by 
domain walls or stripes, 
with holes localized in the nodal regions.

There exists an important difference 
in the physics of the mean-field ground state of the 3D system
and its 2D counterpart. While, in the latter, the system 
remains insulating when
lightly doped, the 3D model immediately turns metallic.
The difference is a consequence of the different behaviors of the modulating wavelength. 
In 2D, $\alpha$ is unity, independent of $h$ and $U$, while in 3D it varies with parameters and has a non-integer value.
To illustrate this in a simple case, consider a value of doping $h$ such that $\lambda$ ($\equiv \alpha/h$) is
an integer. 
For such a system to be an insulator, the number of particles in a $1\times 1\times \lambda$ cell,
$(1-h)\lambda=\lambda-\alpha$, will have to be an integer.
However, 
because $\alpha \sim 2/3$ in the limit of small doping,
the condition 
cannot be satisfied, and the system is necessarily metallic.
A related way to see this is to consider the case of domain wall states, for example
 $U=3.5$ in Fig.~\ref{fig:rho_m_n095_3D}.  Inside each domain wall (nodal region) are localized holes whose integrated (along the direction of the modulation) density is $\alpha$. Thus 
the domain wall as a whole will act as a quasi-2D liquid of holes with non-integer density.
We will discuss the corresponding momentum space signature in the next section.

\subsection{SDW modulation along the $\langle111\rangle$-direction}
\label{ssec:inter_3D}

\begin{figure}
\includegraphics[width=\columnwidth]{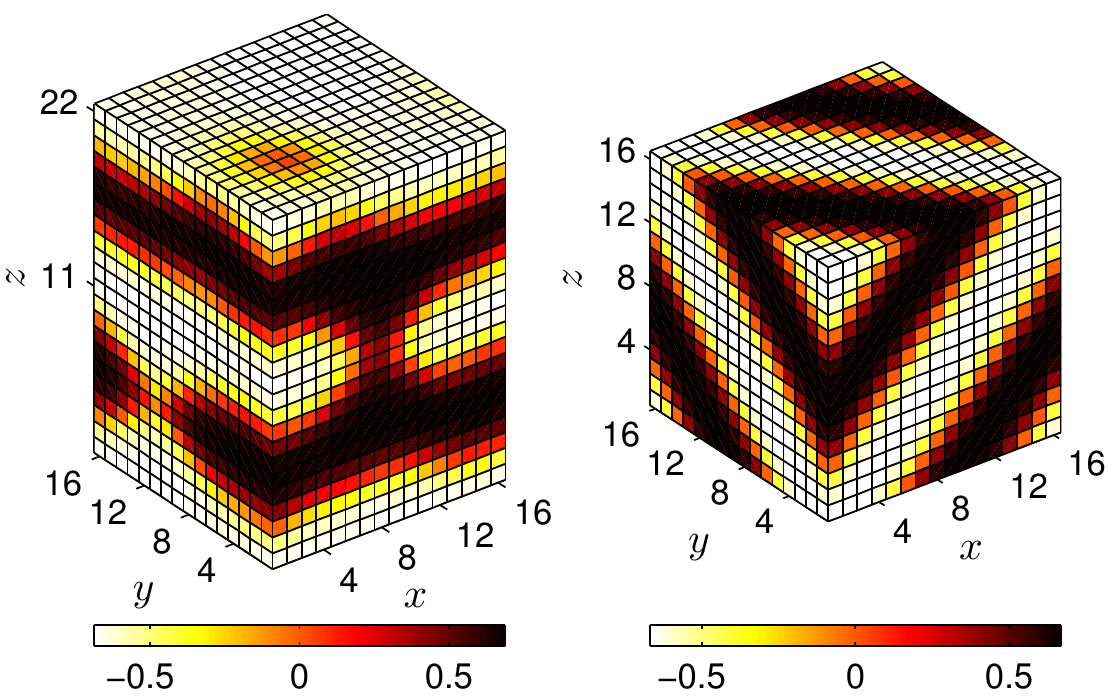}
\caption{
(Color online) Order parameter
$m$ at $h=1/8$ and $U=5.0$
in a $16 \times 16 \times 22$ supercell (left) and a $16 \times 16 \times 16$ supercell (right).
Though the supercell of $16 \times 16 \times 22$ is commensurate
with the optimal $\langle100\rangle$-wavelength,
the random  search produces a lower energy solution.
The minimum energy solution is an SDW along the $\langle111\rangle$-direction,
which is correctly reproduced by a random search
in a $16\times16\times 16$ supercell as shown on the right.
}
\label{fig:m_U50_3D}
\end{figure}

As shown above and further discussed in Sec.~\ref{ssec:pairing},
an orientation of the SDW  
other than $\langle100\rangle$
is not the solution in the proximity of $U_c$.
However, when the interaction grows larger, other Fermi liquid instabilities become possible. 
This fact is clearly
displayed when calculations on supercells commensurate
with the optimal $\langle100\rangle$-wavelength for a given $U$ do not yield a state with $\langle100\rangle$-SDW order. 
Figure~\ref{fig:m_U50_3D} shows the occurrence
of such a case in a calculation with $h=1/8$ and $U=5.0$, for which 
$\lambda_{\langle 100\rangle}=5.5$.
The $16 \times 16 \times 22$ supercell should have precisely accommodated 4 nodal 
planes of the order parameter, but rather than doing so, the random search produces the lower
energy solution shown in the left panel of the figure.

\begin{figure}
\includegraphics[width=\columnwidth]{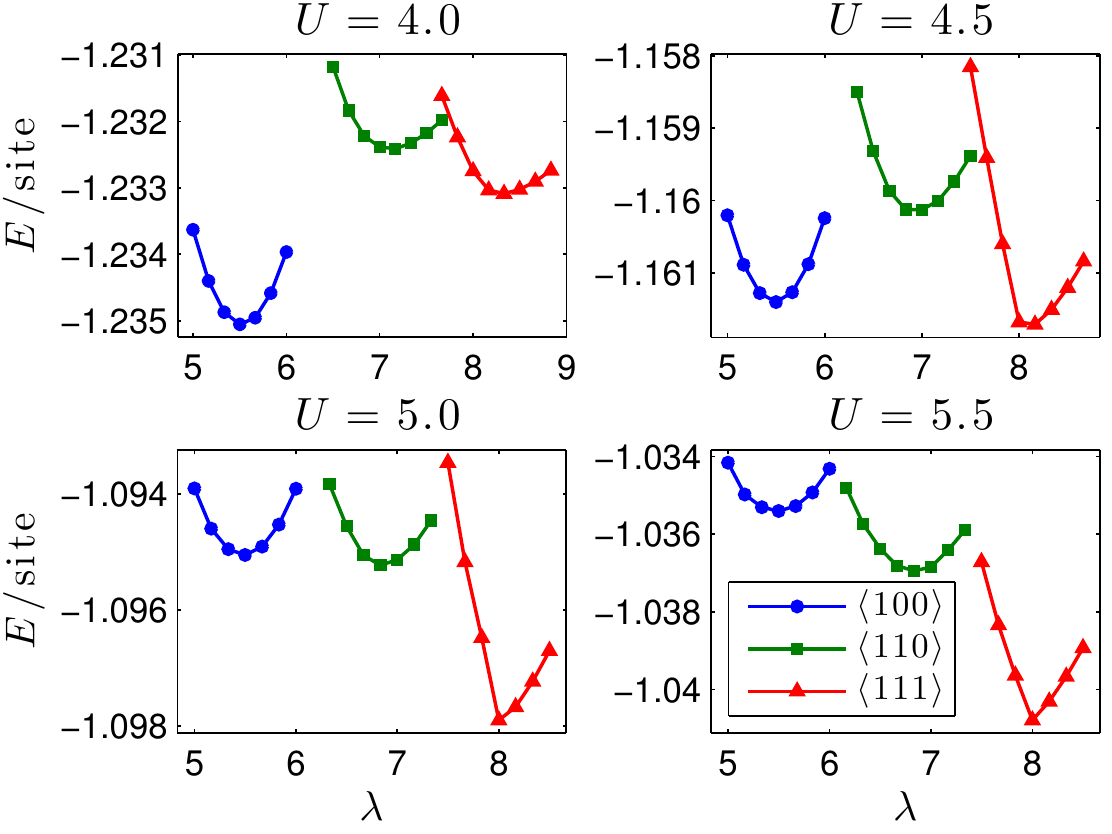}
\caption{
(Color online)
Energies per site of SDW in $\langle100\rangle$-, $\langle110\rangle$- and $\langle111\rangle$-directions
vs. $\lambda$ for $h=1/8$.
At and above $U=4.5$, 
the lowest energy state is modulated along the $\langle111\rangle$- instead of $\langle100\rangle$-direction. 
}
\label{fig:E_n0.875_3D}
\end{figure}

To search for the solution at higher $U$,
we investigate unidirectional SDW's with 
modulation lying along
either the $\langle110\rangle$- or $\langle111\rangle$-direction.
An example is given in Fig.~\ref{fig:E_n0.875_3D}.
The energies from constrained searches using the second approach are shown 
as a function of $\lambda$ for a scan of $U$ values. 
It is seen that, at and above $U=4.5$, the lowest energy state is given by a $\langle111\rangle$-SDW, 
instead of the $\langle100\rangle$-order at lower $U$. For $U=5$, the minimum energy solution is 
correctly reproduced by a random search in a $16\times16\times 16$ supercell
as shown in the right panel of Fig.~\ref{fig:m_U50_3D}.
In our searches, $\langle110\rangle$-direction SDW's are never found to be the global ground state.

\begin{figure}
\includegraphics[width=\columnwidth]{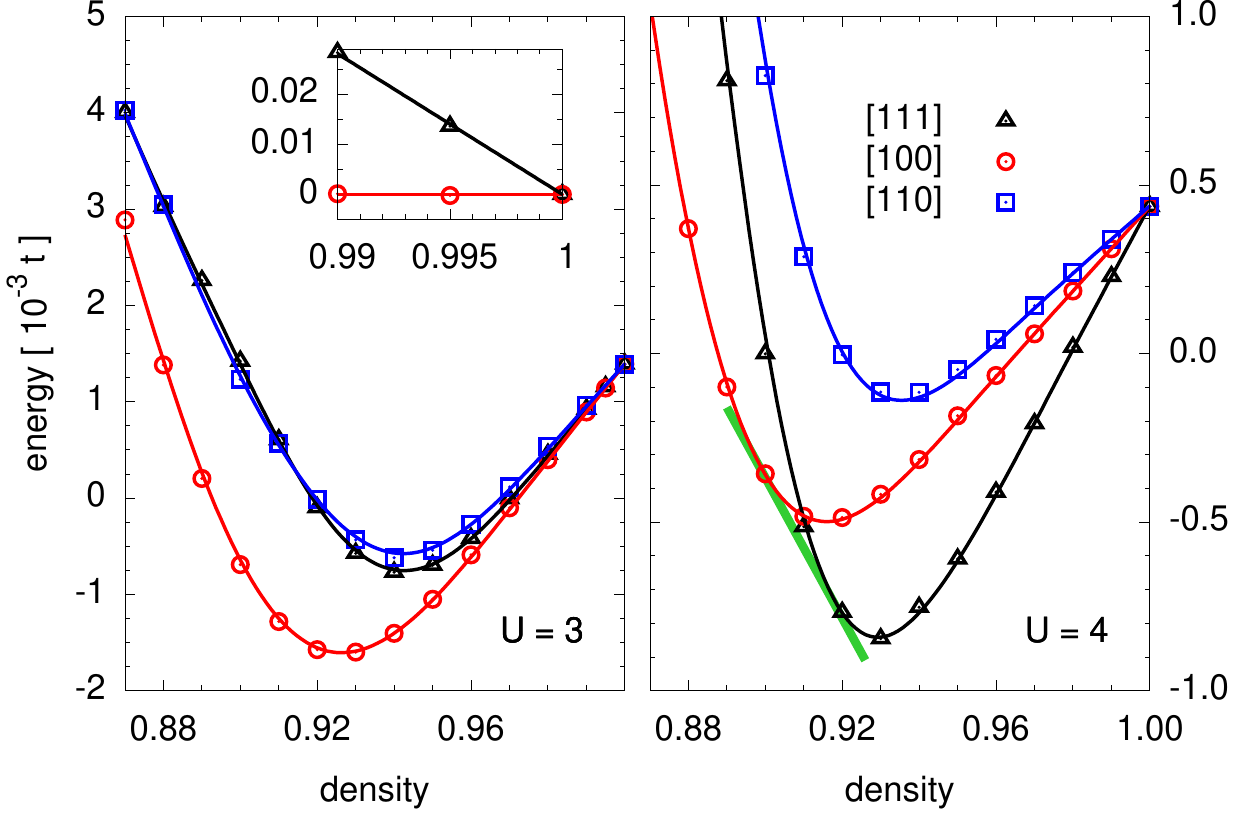}
\caption{
(Color online) Ground-state energy per site from constrained search of 
$\langle100\rangle$-, $\langle110\rangle$- and $\langle111\rangle$-SDW at $U=3$ (left) and $U=4$ (right).
A linear common shift has been applied to the energies to highlight the convexity and the different trends.
At $U=3$,  $\langle111\rangle$-SDW is not the global ground state. (The inset shows a zoom of the 
energy difference between $\langle111\rangle$- and $\langle100\rangle$-SDW states as $n\rightarrow 1$.)
At $U=4$, the ground state is $\langle111\rangle$-SDW for $n \gtrsim 0.92$.
}
\label{fig:U4_vs_U3}
\end{figure}

By repeating the same procedure, we construct the equation of states for $U=3$ and $U=4$ 
contained in Fig.~\ref{fig:U4_vs_U3}. 
At $U=3$, there is no density regime
where the $\langle111\rangle$-SDW is the global ground state.
In contrast, for $U=4$, a discontinuous transition from $\langle100\rangle$ to $\langle111\rangle$
occurs around $n=0.9$ with a small coexistence region.
In both cases the low-doping ground state is characterized by a linear
energy-density dispersion. This bears two important consequences. First,
contrary to what is observed using variational states with a uniform spiral
order parameter \cite{Igoshev2010}, there is no
sign of phase separation into a half-filled, AFM region and a hole-rich region.
Second,
the effective interaction between domain walls is short-ranged and
their precise location in the hole-diluted limit is therefore irrelevant as long as they
stay sufficiently far apart. 

We find $\alpha_{\langle 111 \rangle}=1$ at any density  
for which the $\langle 111 \rangle$-SDW is the 
ground state.
The $\langle 111 \rangle$-SDW states are fully gapped,
owing to the integer value of $\alpha_{\langle111\rangle}$, in 
contrast to the
metallic behavior of the $\langle 100 \rangle$ states. Upon increase
of $U$ at a constant $h$, the structural transition is therefore 
always accompanied by a metal-to-insulator transition. 
We have verified, for selected cases,
that random searches on larger supercells with sizes commensurate to the optimal 
wave-vector always return unidirectional SDW's with the same predicted wavelength and orientation.
This provides a strong indication that the character
of intermediate $U$ instabilities remains that of a unidirectional SDW.
Thus, as we increase $U$ at constant 
density, the system is always expected to undergo a discontinuous transition from 
a $\langle100\rangle$- to a $\langle111\rangle$-SDW ground state.

\subsection{A variational pairing ansatz}
\label{ssec:pairing}

The pairing model is
a variational ansatz that has 
proved extremely helpful in rationalizing
the properties of SDW's in the mean-field treatment of the electron gas\cite{Overhauser, Zhang2008}
and the 2D Hubbard model\cite{Xu2011}. 
Similarly here, the model helps to explain the numerical results and 
provides a simple conceptual framework that captures the essential physics
of the SDW states in 3D and in the crossover regime discussed in the next section.
We first summarize the formalism, and then apply it
to the case of $\langle100\rangle$-SDW at modest $U$, followed by the $\langle111\rangle$-SDW.

At low $U$ and small $h$, the pairing model is defined by
spin-orbitals of the form
\begin{equation}
\phi_{\bfk\sigma}^\dagger= u_\bfk c^\dagger_{\bfk\sigma}+\sigma v_\bfk c^\dagger_{\bfk+\bfq_\bfk\sigma}.
\label{eq:orb_ansatz}
\end{equation}
The construction requires excitations to outside the Fermi sea. 
It is reminiscent of the ansatz used to construct the BCS pairing states for attractive interactions,
except that here the tendency for small 
excitations is perhaps more 
``natural", because of the repulsive interaction.  
A collection of spin pairs in orbitals given by Eq.~(\ref{eq:orb_ansatz}) leads to a uniform charge density,
$\rho(\bfR) = n$, and a spin density of the form
\begin{equation}
s(\bfR) = \frac{4}{N}\sum_{\bfk\in \mathcal{R}} a_\bfk \cos(\bfq_\bfk \cdot \bfR)
\end{equation}
with $a_\bfk=u_\bfk v_\bfk$.
The region 
$\mathcal{R}$ over which $\bfk$ is summed will be closely related to the 
non-interacting Fermi sea, and preserving the volume of $4\pi^3n$, but will
in general be slightly modified from the variational optimization, as we further discuss below.
To ensure orthogonality amongst the spin-orbitals, 
$\bfq_\bfk$ must be such that $\bfk+\bfq_\bfk\not\in \mathcal{R}$ and  $\bfk+\bfq_\bfk\neq \bfk'+\bfq_{\bfk'}$.
The corresponding potential energy per site is then given by
\begin{equation}
V = U n^2 -\frac{U}{4N} \sum_{\bfR} s^2(\bfR).
\label{eq:fullV}
\end{equation}
The potential energy lowering relative to the paramagnetic (PM) solution is thus:
\begin{eqnarray}
\Delta V &=& -\frac{U}{N^2}\sum_{\bfk,\bfk'\in\mathcal{R}} a_\bfk a_{\bfk'} \left[\delta(\bfq_\bfk+\bfq_{\bfk'}) + \delta(\bfq_\bfk - \bfq_{\bfk'})\right], \nonumber \\
&&
\label{eq:dV}
\end{eqnarray}
where the Kronecker $\delta$ is intended as periodic on the reciprocal lattice, {\em i.e.}, modulo $2\pi$
in any direction.

\begin{figure}
\includegraphics[width=\columnwidth]{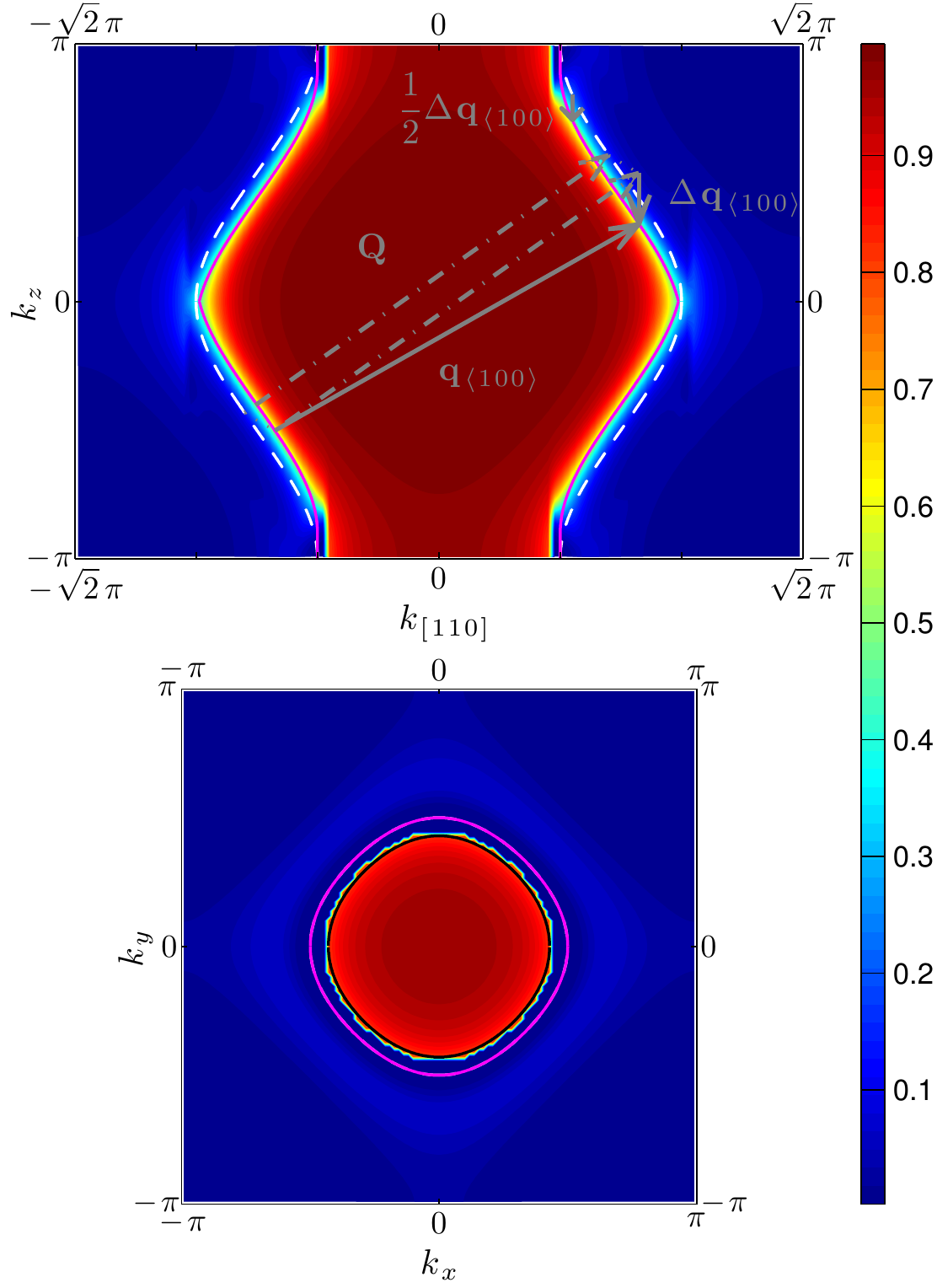}
\caption{
(Color online)
Illustration of the pairing model for the $\langle100\rangle$-SDW state in 3D.
The schematic diagram is drawn on the actual momentum distribution from the exact 
numerical solution for $h=1/8$ and $U=4.0$.
The top panel shows the pairing construction on the contour plot of the $(1\bar10)$ cut.
The white dashed lines represent the half-filled surface,
across which is the nesting vector $\bfQ$.
The reconstructed surface, shown as magenta solid lines, is obtained
by displacing the half-filled one along the $z$-direction
by a distance $\Delta q_{\langle100\rangle}/2$, as given by Eq.~(\ref{eq:dopedV_001}).
The nesting vector across the shifted surface,
$\bfq_{\langle100\rangle}$, is shown by the long solid line with arrow. 
The bottom panel shows $n(\bfk)$ of the $k_z=\pi$ plane. This is in a region 
where the Fermi surface survives the onset of order and where
it differs more severely from the pairing construction.
The actual Fermi surface, seen distinctly inside the reconstructed surface,
differs little from the non-interacting Fermi surface (black solid line). 
$n_{\bfk}$ drops sharply, and no pairing is present in this region. 
}
\label{fig:theory_3D}
\end{figure}

Equation~(\ref{eq:dV})  makes it clear that the maximum reduction in $V$
is achieved by having as many pairs as possible
with
$\bfq_\bfk$'s
which are parallel or anti-parallel to each other.
Noting that the vector $\bfQ$ is perfectly nested when $h=0$, let us
consider the following explicit construction for $\mathcal{R}$:
displace the half-filled Fermi surface in each octant of the first BZ by $\pm \Delta\bfq/2$,
choosing the direction that shrinks the Fermi sea
and with a length $\Delta q$ such that the enclosed volume is reduced to $4\pi^3 n$.
This construction is illustrated in the top panel in Fig.~\ref{fig:theory_3D} for the case
of the $\langle 100\rangle$-SDW (discussed next in Sec.~\ref{ssec:pairing_100}). 
The surface of 
$\mathcal{R}$
can now anchor spin pairs with one common pairing vector,
by making $u_\bfk$ less than $1$ in a small layer immediately inside the surface of 
 $\mathcal{R}$, and correspondingly $v_\bfk=\sqrt{1-|u_\bfk|^2}>0$.

For small $h$,  we can determine $\Delta q$ directly from the construction:
\begin{equation}
\Delta q\int\limits_{\rm S} \widehat{\bfe}_{\Delta \bfq} \cdot{\rm d}{\bfS} = h\frac{\Omega_{\rm BZ}}{2},
\label{eq:dopedV_001}
\end{equation}
where S is the half-filled surface,  $\widehat{\bfe}_{\Delta \bfq}$
is the direction of $\Delta \bfq$, and $\Omega_{\rm BZ}=(2\pi)^3$ is the volume of the first
BZ.
Equation~(\ref{eq:dopedV_001}) implies a linear relationship between $h$ and $\Delta q$
and, using $\lambda = \pi/(\Delta \bfq \cdot \widehat{\bfe})$, provides the following estimate of $\alpha$
(Eq.~(\ref{eq:def_alpha_CDW})) 
\begin{equation}
\alpha_{\Delta \bfq}=\frac{1}{4\pi^2~\widehat{\bfe}_{\Delta \bfq} \cdot \widehat{\bfe}}\int\limits_{\rm S} \widehat{\bfe}_{\Delta \bfq} \cdot{\rm d}{\bfS}.
\label{eq:alpha}
\end{equation}
where $\widehat{\bfe}$ is a relevant Cartesian unit vector.

Different directions of $\Delta \bfq$ lead to 
different 
reconstructions of the non-interacting doped Fermi surface. The kinetic energy 
cost of the pairing ansatz can therefore be thought of as the combination of two contributions: the reconstruction 
energy due to using $\mathcal{R}$ rather than the true non-interacting Fermi sea, and the 
kinetic energy change due to moving particles from $\bfk$ (inside $\mathcal{R}$) to 
$\bfk+\bfq_\bfk$ (outside), {\em i.e.}, the non-zero $v_\bfk$'s.
It is easy to see that, similar to 2D\cite{Xu2011}, 
at sufficiently large $U$ the potential energy lowering  will overtake the kinetic energy increase
for the constructions discussed here.
The correct state is determined by 
maximizing the gain in the potential energy from pairing
(larger areas near the Fermi surface participating with parallel $\Delta \bfq$)
while minimizing the kinetic energy cost. 

The ansatz gives a clear picture for the
onset of the instability. First, by its form, the model captures how 
the energetically costly
CDW can be suppressed compared to the SDW.
Second, it indicates that amongst different possible
choices of $\bfq_\bfk$, the one involving only parallel and anti-parallel vectors
are optimal. Third, the direction of $\pm \Delta \bfq$ must be such as to lead to
the minimal possible reconstruction of the doped Fermi surface. 
These findings are in qualitative agreement with the numerical results obtained by the solution of the SCF
equations:
1) the SDW is much stronger than the accompanying CDW,
 2) the Fermi surface reconstructs  in
a way to enhance pairing and the SDW order
tends to be unidirectional as a result,
 3) more drastic reconstructions are
only possible
with larger $U$.
Much quantitative information can be obtained with straightforward calculations using this model,
as we discuss next for the $\langle100\rangle$- and $\langle111\rangle$-SDW states,
respectively.

\subsubsection{Analysis of the $\langle100\rangle$-SDW}
\label{ssec:pairing_100}

Among all directions, only a ${\Delta \bfq}$ along the $\langle100\rangle$-direction 
causes the Fermi surface in each octant to shrink equally and this, it can be shown,
leads to the minimum kinetic energy increase at small $h$. 
An SDW with $\langle100\rangle$ modulation is thus the lowest energy solution at low $h$ and just above $U_c$,
consistent with the results from explicit solutions of the SCF equations in Sec.~\ref{ssec:weak_3D}.
We numerically calculate the projected area along the $\langle100\rangle$-direction
(shown in the right middle panel in Fig.~\ref{fig:halffillingFS})
and obtain
$\alpha_{\langle100\rangle}\simeq0.63$,
in very good agreement with the exact results from direct solutions shown in Fig.~\ref{fig:alpha_3D},
where $\alpha_{\langle100\rangle} \simeq$ 0.66.
That the estimated value is slightly smaller is consistent with the presence of surviving
Fermi surface inside the reconstructed doped Fermi surface as seen in Fig.~\ref{fig:theory_3D}.

A direct comparison between the pairing model and the exact SCF solution 
can also be made in momentum space. 
We will identify the 
Fermi surface in the numerical solution from mean-field theory 
as the locus of points where $n_{\bfk \sigma} =0.5$.
Depending on the system and the value of $U$, the 
momentum distribution of the exact mean-field ground state can
 maintain a true Fermi surface, characterized by a discontinuity in $n_\bfk$, or
have it smeared out by large pairing amplitudes ({\em i.e.}, $u_\bfk$ close to 
$1/\sqrt{2}$ near the boundary of $\mathcal{R}$).
The two scenarios can occur in the  same system at different $\bfk$ values.
The identification using $n_{\bfk \sigma} =0.5$ is 
consistent with both.

\begin{figure}
\includegraphics[width=\columnwidth]{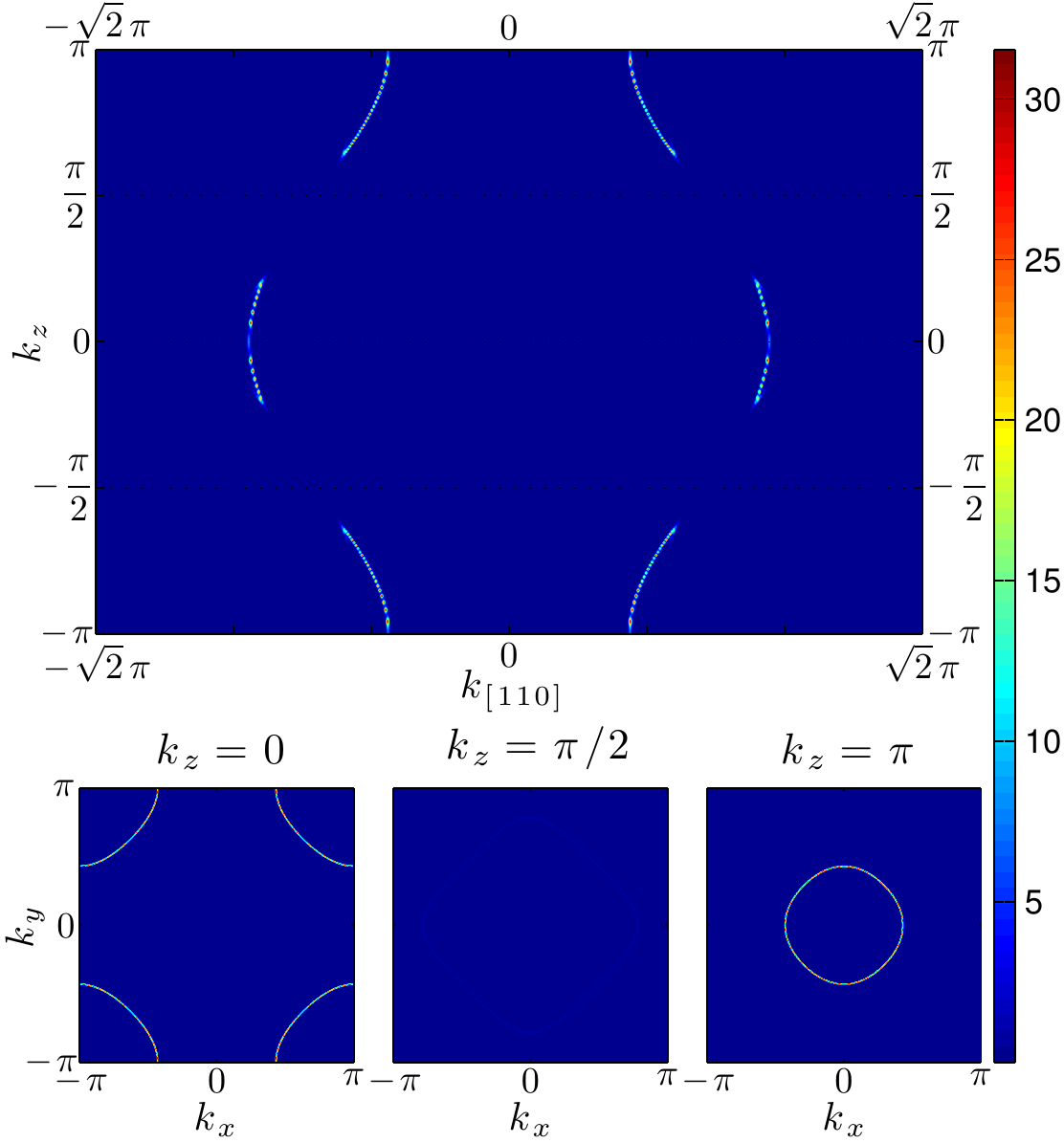}
\caption{
(Color online)
Contour plots of the single-particle spectral function evaluated
at the Fermi energy;
on the $(1\bar10)$ cut (top) and on the $k_z=0,\pi/2,\pi$ planes (bottom from left to right)
for a system with $U=2.7$ and $h=1/8$.
A large part of the Fermi surface survives, except for areas around the hot spots
where it is most energetically favorable for pairing. 
}
\label{fig:fs_3D}
\end{figure}

Figure \ref{fig:theory_3D} shows, in particular, that the portion of the Fermi surface where pairing takes place is in very good agreement 
with the construction based on the pairing model, which indicates that the ansatz captures 
the dominant ingredient of the physics of the SDW state.
The figure also provides a direct
explanation for the survival of the Fermi surface around $k_z=\pi$ as it is
there that the pairing construction shows large discrepancy with the true Fermi surface.
This, in turn, implies that pairing in that region would be associated with too large a kinetic energy cost to be 
favorable.
The absence of any gap from pairing at the Fermi surface
in the $k_z=\pi$ plane is consistent with the picture discussed earlier of a quasi-2D liquid within each domain wall.
These effects are amplified at smaller $U$'s
as shown in Fig.~\ref{fig:fs_3D}, where a larger part of the 
Fermi surface survives the onset of order; the parts that do not survive are
in areas around the ``hot spots'' $\bfk\simeq(\pm\pi/2,\pm\pi/2,\pm\pi/2)$,
where the pairing construction
and the doped Fermi surface are most similar, 
and the change in $\frac{\partial \epsilon_{\mathbf k}}{\partial k_z}$
is at a minimum implying a closer proximity to a perfect common paring vector.

\begin{figure}
\includegraphics[width=\columnwidth]{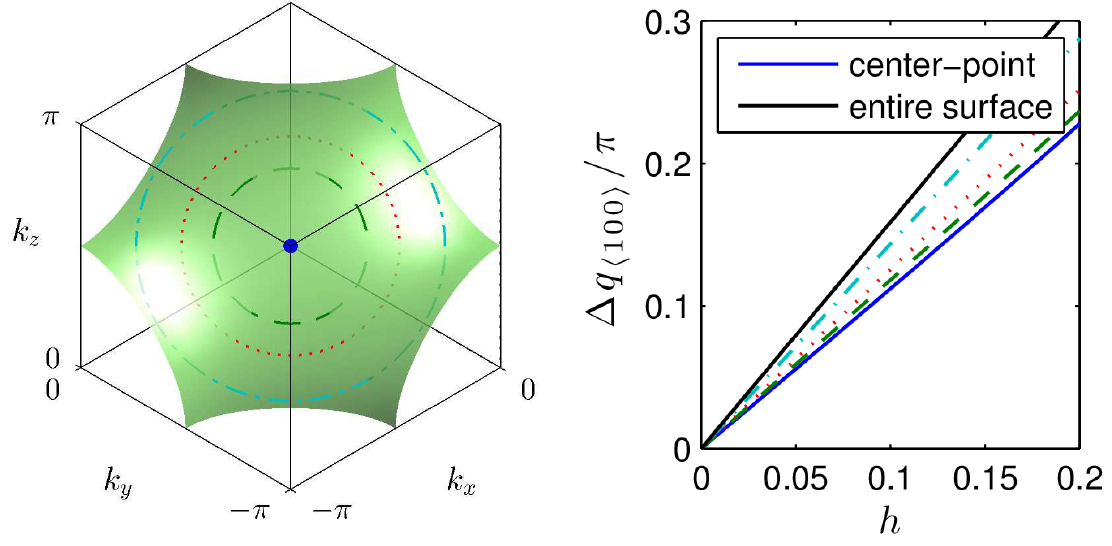}
\caption{
(Color online)
(a) Half-filled Fermi surface for the 3D Hubbard model in one of
the octants (left) and (b) averaged $\Delta q_{\langle100\rangle}$
over different areas vs.~doping (right).
The blue solid line in (b) is calculated
at $(-\pi/2,-\pi/2,\pi/2)$, shown as the blue dot in (a).
The dashed/dotted lines in (b) are averaged
over varying areas as indicated by the circles with the same color/style in (a). 
The black line in (b) shows the averaged value over the entire surface, 
which
leads to the estimate of $\alpha_{\langle100\rangle}\sim 0.63$ discussed in the text.
}
\label{fig:d_3D}
\end{figure}

These understandings allow quantitative explanations of all the features of the 
data on $\alpha_{\langle100\rangle}$ shown in Fig.~\ref{fig:alpha_3D}. 
For example, 
Fig.~\ref{fig:d_3D} shows that the distance along $z$-direction between the doped
and half-filled Fermi surfaces is at a minimum around the hot spots.
Given that such distance equals $\Delta q/2$ in the pairing construction,
one finds that the local $\Delta q$ value is smaller at the hot spots
than when computed as the average distance over the entire surface. Hence, when only the hot spots are
involved in pairing, a larger $\alpha_{\langle100\rangle}$ results,
as seen at lower values of $U$ just above $U_c$.
Obviously, the pairing reconstruction becomes increasingly
accurate as $h$ approaches 0. In this limit, one therefore finds
the increasingly smaller $U_c$ and the faster convergence (in $U$) to the
saturated value of $\alpha_{\langle100\rangle} \sim 2/3$ shown in Fig.~\ref{fig:alpha_3D}.

\subsubsection{Application to the $\langle111\rangle$-SDW}
\label{ssec:pairing_111}

\begin{figure}
\includegraphics[width=\columnwidth]{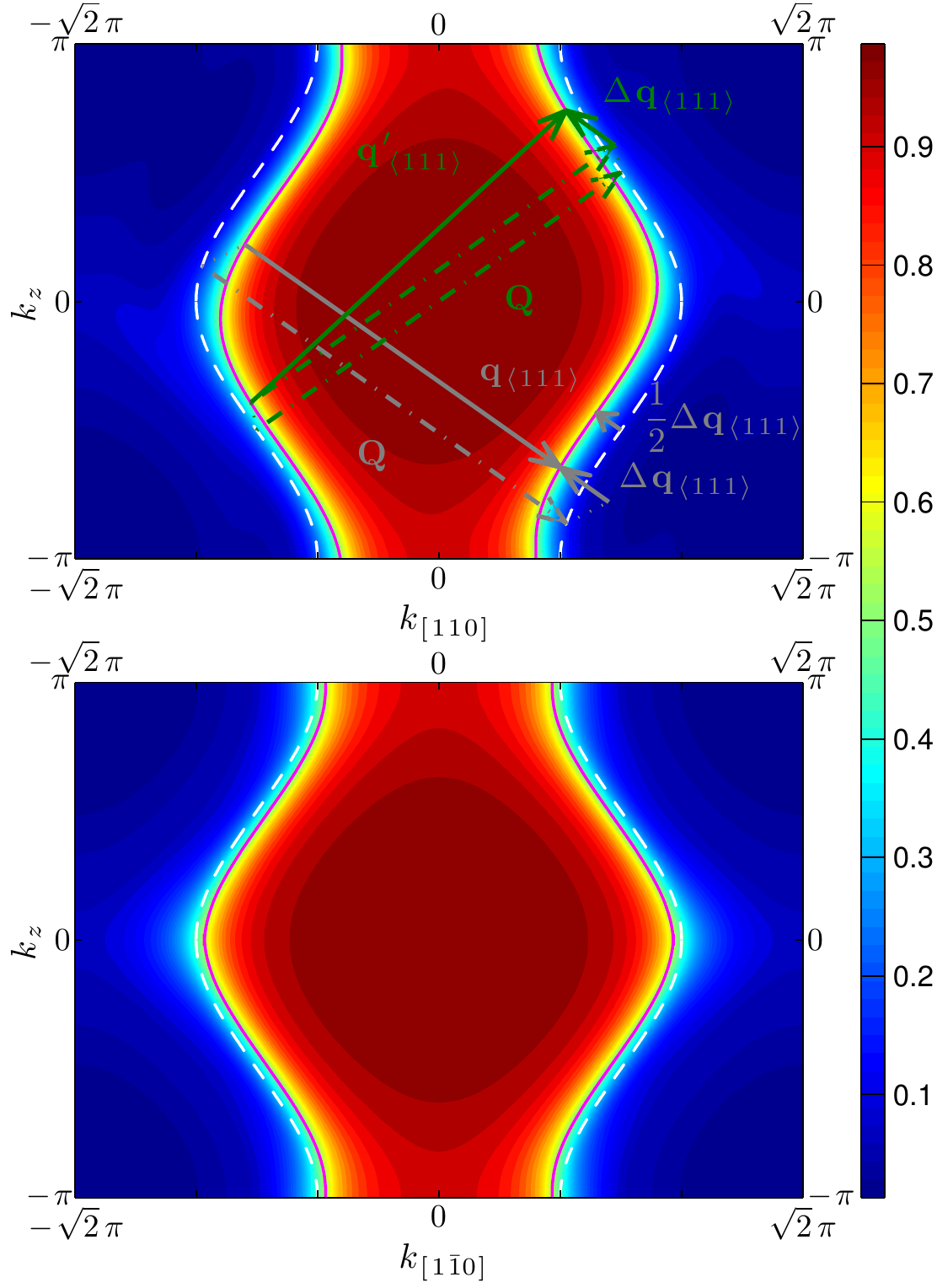}
\caption{
(Color online)
Schematic illustration of the pairing model for the $\langle111\rangle$-SDW state,
shown on the contour plots of the $(1\bar10)$ (top) and the $(110)$ (bottom) cuts
of
$n_{\bfk \uparrow}$
for a system of $h=1/8$ and $U=5.0$.
The white dashed lines are the half-filled surface,
across which is the nesting vector $\bfQ$.
The magenta solid lines show the pairing construction, obtained
by shifting the half-filled surface along $[\bar1\bar11]$-direction
by a distance of $\Delta q_{\langle111\rangle}/2$.
In the upper panel, $\bfq_{\langle111\rangle}$ and $\bfq'_{\langle111\rangle}$ give two 
equivalent representations (differing by a reciprocal lattice vector) of the pairing 
vector across the reconstructed Fermi surface. 
Note the asymmetry between the two diagonal directions in the upper panel.
}
\label{fig:k_1113D}
\end{figure}

By taking a displacement $\Delta \bfq$ along the $\langle111\rangle$-direction, 
we can apply the pairing construction straightforwardly to the diagonal-modulated SDW.
Analogously to the $\langle 100\rangle$ case, 
Fig.~\ref{fig:k_1113D} shows a remarkable agreement
between the shifted half-filled surface and the 
calculated Fermi surface of the SDW.
The two cuts in the figure clearly show broken cubic symmetry,
where the Fermi
surface in one pair of the octants is further away from the half-filled one
so as to share the common modulation wave-vector $\Delta \bfq_{\langle111\rangle}$ with the other three pairs.
The $\langle111\rangle$-SDW state offers, in this respect, a particularly clear example
where Fermi surface reconstruction can be observed. 
It also shows how
 an accurate experimental
characterization of the momentum distribution in optical lattices
can be used to characterize the band structure and pairing at the Fermi surface which, in turn,
provides momentum space evidence on the real space character of the SDW.

Using Eq.~(\ref{eq:alpha}) we have estimated
the wavelength of the $\langle111\rangle$-SDW, and found 
$\alpha_{\langle111\rangle}=0.93$.
The exact SCF calculations in Sec.~\ref{ssec:inter_3D} showed, instead, that
$\alpha_{\langle111\rangle}$
is precisely pinned at 1 in a fairly large regime of $U$.
This somewhat large discrepancy is a consequence
of the natural tendency of the system to ``lock'' 
the integrated density of holes per nodal region to 1 whenever the topology of the non-interacting doped Fermi surface
is
such that this is not energetically too costly.
When doing so the system benefits from both pairing and 
band-insulating mechanisms to gap the entire Fermi surface and
further lower the ground-state energy.

\section{Dimensional Crossover Results}
\label{sec:crossover}

\subsection{Results from full numerical HF solutions}
\label{ssec:crossover_results}

The mean-field ground state of the doped 2D Hubbard model shares many similarities
with its 3D counterpart. Just above $U_c$,
the 2D system develops a sinusoidal SDW with a modulating wave along the $\langle10\rangle$-direction and a much weaker accompanying CDW.
As $U$ is increased, the SDW increases its amplitude before 
the SDW state eventually changes into a collection of weakly interacting domain walls.
Above a certain $U$,
there is a discontinuous transition to a phase
where the modulation is  along the $\langle11\rangle$-direction. The crossover from 
SDW to domain walls occurs before the $\langle10\rangle$ to $\langle11\rangle$ transition at small $h$, but after at larger $h$ \cite{Xu2011}.
A peculiarity of the 2D case, due to the special topology of the 2D half-filled surface,
is that $\alpha=1$ and the system is an insulator regardless of doping, $U$ or direction 
of the modulation wave-vector apart from a region close to $U_c$.

By controlling the distance between square lattice layers, optical lattice experiments
allow the study of the evolution of the system as it
crosses over from 2D to 3D.  This situation is theoretically described by an increase of
$t_\perp$ in Hamiltonian (\ref{eq:def_ham}) and the question, within mean-field theory,
concerns the ensuing evolution of the ground-state properties.
The pairing model and the arguments described in Sec.~IV.C
remain valid in the crossover regime. We thus restrict our investigation to 
unidirectional SDW ground states, although we did use the first approach to carry out some searches, finding no 
additional structures.
As in 3D, we verify that 
the SDW solution with minimum energy, identified using the second approach, can be obtained
by the first approach in a large supercell that is commensurate with the optimal wavelengths, 
even when starting from random initial guesses. 
SDW in directions different from $\langle100\rangle$ or $\langle111\rangle$
are not found to be the global ground state for any value of $t_\perp$.

\begin{figure}
\includegraphics[width=\columnwidth]{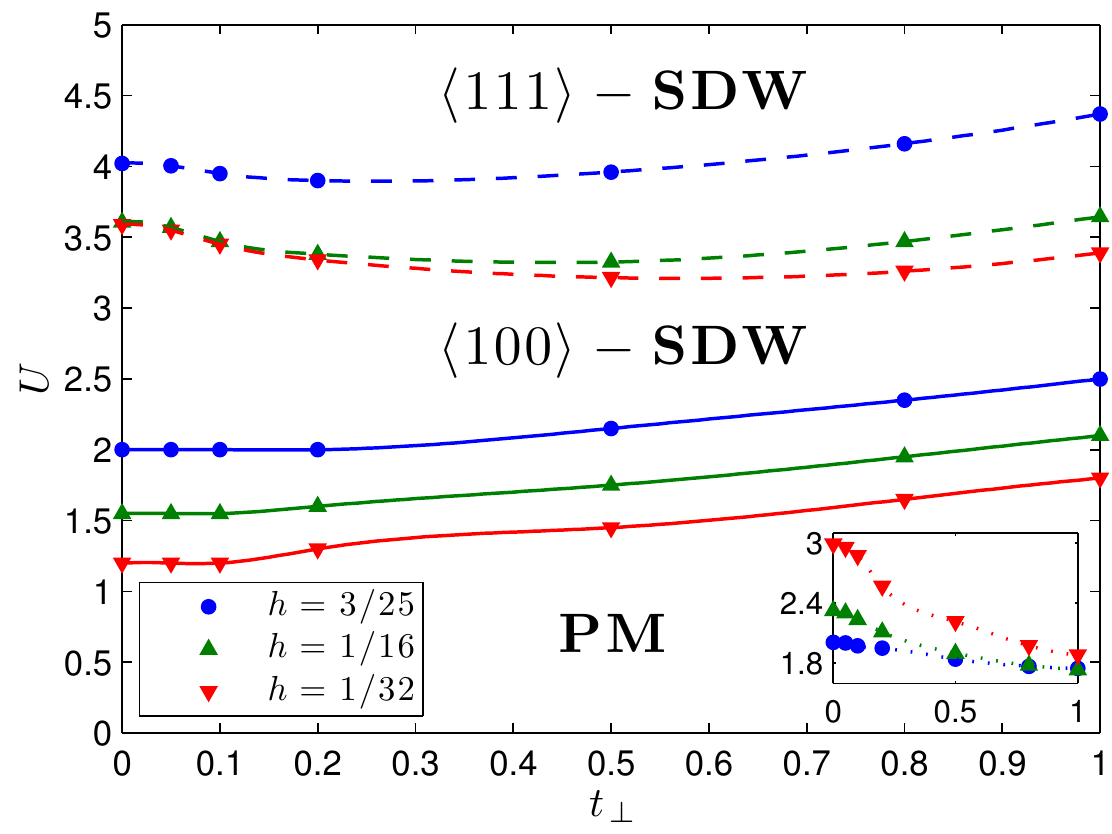}
\caption{ 
(Color online) Mean-field phase diagram of the ground state in the crossover regime.
Phase boundaries for several values of doping are indicated by symbols. The 
lines are to guide the eye. 
Solid lines separate the PM phase from the AFM phase,
and dashed lines show transitions from $\langle100\rangle$- to $\langle111\rangle$-SDW.
The inset plots the value of
$U_{\langle100\rangle \rightarrow \langle111\rangle}/U_{{\rm PM} \rightarrow \langle100\rangle}$.
}
\label{fig:phasediag}
\end{figure}

Results are summarized in the $t_\perp$-$U$ mean-field phase diagram of 
Fig.~\ref{fig:phasediag}. 
An overall increase in the critical $U$ values is seen as doping increases,
as a result of a greater deformation from the perfectly nested half-filled Fermi surface and the need 
for more excitations to achieve reconstruction of the Fermi surface for pairing.
As before, numerical calculations focus on small doping ($h \leqslant 0.2$)
and low to intermediate interactions ($U \leqslant 5.5$), where mean-field
theory can be expected to be more accurate.
Upon increase of $U$, and similarly to 3D, the system undergoes a first transition to a 
$\langle100\rangle$-SDW state followed by a second, discontinuous transition to a
$\langle111\rangle$-SDW state. 
The absence of cubic symmetry away from $t_\perp =1$ causes the modulation wave-vector for 
$\langle100\rangle$-SDW to lie in the $xy$-plane. This is because the elongation of the Fermi surface along $z$-direction
(as illustrated in  Fig.~\ref{fig:halffillingFS}), meaning that $\Delta \bfq$ along $z$-direction leads to less surface area for pairing than along $x$- or $y$-directions.
Wave-vectors along the $\langle111\rangle$-direction continue, on the other hand, to remain equivalent under
the symmetry operation of the tetragonal group.

The critical value of the interaction strength for the transition from the paramagnetic (PM) phase to 
$\langle100\rangle$-SDW, $U_{{\rm PM} \rightarrow \langle100\rangle}$,
monotonically increases from 2D to 3D, due to the wider band width and smaller density of states at
the Fermi energy for larger $t_\perp$. 
The transition values decrease to 0 when $h$ approaches 0
as $U_c=0$ for the half-filled system at any $t_\perp$.
The critical $U$ value for the transition from $\langle100\rangle$- to $\langle111\rangle$-SDW,
$U_{\langle100\rangle \rightarrow \langle111\rangle}$,
has a lower bound lying close to the $h=1/32$ line in
the figure, so that no $\langle111\rangle$-SDW exists below $U\simeq3$ regardless of the smallness of 
$h$ and the value of $t_\perp$. In contrast with $U_{{\rm PM} \rightarrow \langle100\rangle}$,
$U_{\langle100\rangle \rightarrow \langle111\rangle}$ 
displays a non-monotonic behavior with $t_\perp$ whose origin we will address in the
next section.

\begin{figure}
\includegraphics[width=\columnwidth]{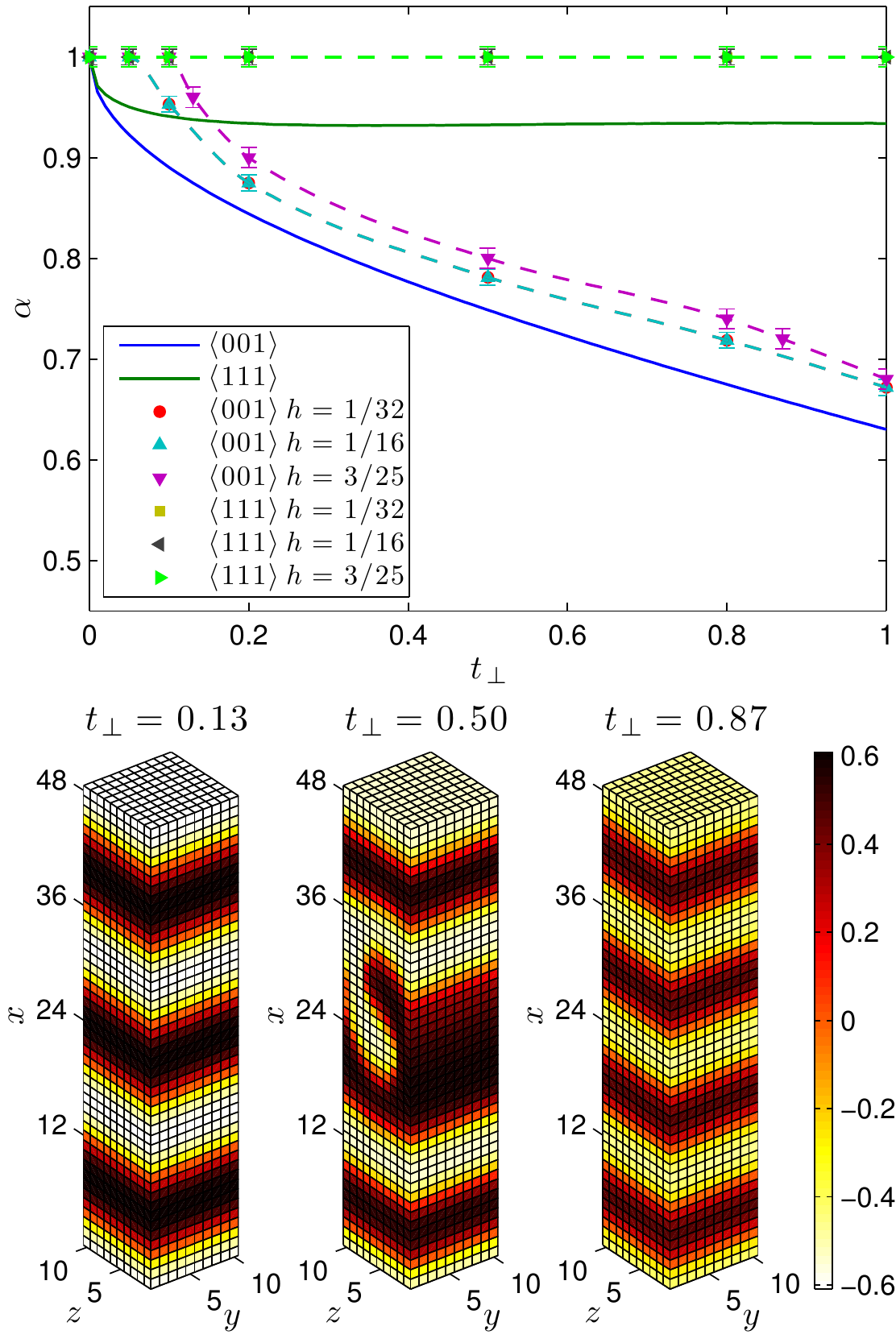}
\caption{
(Color online) Top panel: dependence of the modulation wavelength on $t_\perp$.
Numerical results for $\alpha$ (data points with error bars) at various dopings are
compared with theoretical estimates (solid lines).
Bottom panel: evolution of the order parameter on a $48\times 10 \times 10$ supercell for
$h = 3/25$, $U = 3.5$ as a function of $t_\perp$.
As $t_\perp$ increases, $\alpha_{\langle 100 \rangle}$ remains at 1 in a 
very small region close to 2D, before decreasing monotonically,
while $\alpha_{\langle111\rangle}$ is locked at 1.
}
\label{fig:alpha_co}
\end{figure}

The evolution of the modulation wavelength 
is summarized in Fig.~\ref{fig:alpha_co} in terms of $\alpha$ as a function of $t_\perp$.
Numerical results 
are obtained at $U$ values around the transition line in Fig.~\ref{fig:phasediag}, {\em e.g.}, just below $U_{\langle100\rangle \rightarrow \langle111\rangle}$
for $\langle 100\rangle$. In  a small window around $t_\perp=0$, $\alpha_{\langle 100\rangle}$ 
remains 1, as  a direct consequence of the insulating
character of the 2D solution. Apart from this region, 
there is a gradual decrease of $\alpha_{\langle 100\rangle}$ with $t_\perp$ (discussed further 
in the next section) and a weakening of the intensity of the order parameter (bottom panel of
Fig.~\ref{fig:alpha_co}). As we have already remarked in the 3D results,
the pinning of $\alpha$ at 1 in the $ \langle111\rangle$ phase, and the corresponding insulating behavior, are a result
of a further stabilization of the ground state caused by coexisting magnetic and
band-insulating effects.

\subsection{Pairing model discussion}
\label{ssec:crossover_model}

The dependence of $\alpha$ on $t_\perp$ is
captured by Eq.~(\ref{eq:alpha}), as
the projection of the Fermi surface S($t_\perp$) along $\Delta \bfq$.
The middle and bottom rows of Fig.~\ref{fig:halffillingFS}
show how the projected surfaces, along the $\langle100\rangle$- and $\langle111\rangle$-directions 
respectively, shrink as $t_\perp$ is increased. Results from explicit calculations using Eq.~(\ref{eq:alpha})
are shown as continuous lines in Fig.~\ref{fig:alpha_co}. They display a correct trend but a consistently underestimation of wavelengths.
The quantitative disagreement is  not surprising. 
The most significant reason behind it is 
the tendency of $\alpha$ to be locked at 1, on which
we have already commented and which apparently involves more global considerations than
contained in the pairing model.
The smaller discrepancy for the $\langle100\rangle$-SDW outside the immediate vicinity of 2D,
which increases for higher $h$, is due to the surviving FS
that remains inside the reconstructed doped FS, as we have already remarked in the 3D results. 

We next address the origin of the non-monotonic behavior of
$U_{\langle 100\rangle \rightarrow \langle111\rangle}$. This is the result
of two competing factors. On the one hand, 
the increasing band width with dimensionality leads to an increase of $U_c$,
as demonstrated in the monotonicity of $U_{{\rm PM} \rightarrow \langle100\rangle}$.
On the other hand, 
the geometrical properties of the Fermi surface are such that
the angle between $\Delta \bfq_{\langle111\rangle}$
and the Fermi surface in some of the octants is small when $t_\perp$ is small.
This means that  the displaced Fermi surfaces in those octants will remain close 
to the half-filling counterparts in the construction of $\mathcal{R}$ (Sec.~\ref{ssec:pairing}). As a result,
the other components of the Fermi surface must be displaced further to preserve total volume, 
causing a
more uneven reconstruction
which requires more excitations from the non-interacting Fermi sea, hence larger $U_c$. 
The 2D case offers an extreme example of this as it is characterized by a large fraction of the 
reconstructed doped surface remaining exactly pinned on the half-filled one\cite{Xu2011}. 
To separate this factor from that of the band width, we examine
$U_{\langle100\rangle \rightarrow \langle111\rangle}/U_{{\rm PM} \rightarrow \langle100\rangle}$, which
is plotted in the inset of Fig.~\ref{fig:phasediag}.
A monotonic decrease is seen with $t_\perp$.
Therefore, $U_{\langle100\rangle \rightarrow \langle111\rangle}$
first decreases and then increases as $t_\perp$ goes from 0 to 1.

\section{Summary and discussion}
\label{sec:disc&conclu}

This work addressed quantitative aspects of possible inhomogeneous magnetic phases
of the 3D Hubbard model that emerge as the average
density deviates from one particle per site. 
Because of the ease
with which experiments are expected to be able to transition between the
2D and 3D regimes, we also studied the evolution of the
inhomogeneous ground state as a function of the hybridization between
parallel layers of square lattices.
Within mean-field theory, we have shown that the
leading instability of the PM ground state is an SDW 
with long wavelength modulation along the $\langle100\rangle$-direction.
No tendency toward phase
separation was seen, even at small values of doping. The system remains
metallic, regardless of the proximity to half-filling, because of
a non-integer density of holes per wavelength of modulation. This
density is largely determined by an entirely geometric property of the Fermi surface:
its projected area along the direction of modulation.
At larger $U$ values,  the ground state
continues to be a unidirectional SDW, but with $\langle111\rangle$-orientation.
This phase is insulating and characterized by a
significant distortion of the momentum distribution. Such distortion
leads, quite naturally, to the identification of a reconstructed
Fermi surface whose observation in optical lattice experiments
should be feasible. 

We showed that much of these results can be understood
by a simple variational ansatz with pairing orbitals formed by
a linear combination of two plane waves. By placing a pair of up- and down-spin 
particles into a pair of such orbitals, an SDW is formed with constant charge density. 
Straightforward  analysis of the energetics from this ansatz leads to quantitative predictions
of the wavelength and nature of the SDW modulation which are verified by our 
direct numerical solutions of the SCF equations.

The true many-body ground state will modify the mean-field solutions
in several ways.
For example, quantum Monte Carlo calculations in periodic simulation cells will restore translational invariance, and the inhomogeneities seen here will be manifested in spin-spin correlations and such. A deeper issue is the possible existence of additional competing instabilities once a fuller treatment of quantum fluctuations is included. Certainly, the tendency for magnetic inhomogeneous order is exaggerated in mean-field theory,
and a more accurate description of the many-body correlation at a certain $U$ value tends to be given by the mean-field results at a significantly weaker $U$. However, as we have shown in 2D, mean-field theory appears to capture the correct basic picture of the magnetic correlations when compared to quantum Monte Carlo results\cite{Xu2011,Chang2010}. This indicates that the results in the present paper can provide a useful framework for understanding the magnetic correlations in 3D and in the crossover regime for weak to intermediate interaction strengths.

Apart from the obvious omissions inherent in the mean-field approximation,
this study has not addressed the fact that experiments are performed in the presence of
a confining potential.
Nor have we addressed how the situation is
modified by a finite magnetization. 
Generalization of the present approach to address such issues will be valuable, and 
technically straightforward.

\begin{acknowledgments}
The work was supported by NSF (DMR-1006217) and ARO (56693-PH).
Computational support was provided by the Center for Piezoelectrics by Design.
We thank R.~Hulet, H.~Krakauer, and E.~Rossi
 for useful discussions.
\end{acknowledgments}

\end{document}